\documentclass[12pt]{book}
 \usepackage{amsmath,amssymb,amsbsy,amstext,amscd,amsfonts}
 \usepackage{exscale,calrsfs,times,multind,theorem}
 \usepackage{color}
 \usepackage{article}
 \usepackage{pictex,epsfig}
 \input article.def
\begin{document}

\vskip 2.3truecm
\begin{center}{\bf\Large LAPLACE-BELTRAMI EQUATION ON HYPERSURFACES AND $\Gamma$-CONVERGENCE}
\vskip 0.7truecm
{\normalsize  Tengiz BUCHUKURI, Roland  DUDUCHAVA \& George TEPHNADZE }\footnote{The investigation is supported by the grant of the Shota Rustaveli Georgian National Science Foundation GNSF/DI/10/5-101/12}

\thispagestyle{empty}
\vskip 0.7truecm
\baselineskip=12pt
\thispagestyle{empty}
 \vskip1mm
 \end{center}

\vskip 0.25in

\begin{quote}
{\bf Abstract.} We investigate a mixed boundary value problem for the stationary heat transfer equation in a thin layer with a mid hypersurface $\cC$ in $\bR^3$ with the boundary. The main object is to trace what happens in $\Gamma$-limit when the thickness of the layer converges to zero. The limit Dirichlet BVP for the Laplace-Beltrami equation on the surface is described explicitly and we show how the Neumann boundary conditions in the initial BVP transform in the $\Gamma$-limit. For this we apply the variational formulation and the calculus of G\"unter's tangential differential operators on a hypersurface and layers, which allow global representation of basic differential operators and of corresponding boundary value problems in terms of the standard Euclidean coordinates of the ambient space $\mathbb{R}^n$.
\end{quote}

\tableofcontents
%

 %
 %
\section*{Introduction}
\label{sec0}
\addcontentsline{toc}{section}{Introduction}
\setcounter{equation}{0}

The main aim of this paper is to demonstrate what happens with a boundary value problem for the Helmholtz equation in a thin layer $\Omega^\varepsilon$ with a mid hypersurface $\cC$ in $\bR^3$ when the thickness of the layer $2\varepsilon$ diminishes to zero $\varepsilon\to0$. We impose the Neumann boundary conditions on the upper and lower faces of the layer $\mathcal{C}\times\{\pm\varepsilon\}$ and  the Dirichlet boundary conditions on the
lateral surface $\partial\mathcal{C}\times(-\varepsilon,\varepsilon)$.

The convergence is understood in the $\Gamma$-convergence sense. Equation in the layer is represented in terms of the extended Gunter's derivatives-the system of tangential Gunter's derivatives on the surface. the column of surface gradient
 \begin{equation}\label{eq1.1}
 \mathcal{D}:=(\mathcal{D}_1,\mathcal{D}_2,\mathcal{D}_3)^\top
 \end{equation}
(cf. \cite{Gu1}, \cite{KGBB1}, \cite{Du1}). Here $\cD_j:=\pa_j-\nu_j\pa_\nub$ is the G\"unter's tangential derivative on the mid surface $\cC$ and $\nub=(\nu_1,\ nu_2,\nu_3)^\top$ is the unit normal vector field on $\cC$. The first-order differential operator $\mathcal{D}_j$ is the directional derivative along $\pi\,e_j$, where $\pi:\mathbb{R}^n\to T\mathcal{C}$ is the orthogonal projection onto the tangent plane to $\mathcal{C}$ and, as usual, $e_j=(\delta_{jk})_{1\leq k\leq n}\in\mathbb{R}^n$, with $\delta_{jk}$ denoting the Kronecker symbol.

Calculus of Gunter's derivatives on a hypersurface allows representation of the most basic partial differential operators (PDO's), as well as their associated boundary value problems, on a hypersurface $\mathcal{C}$ in $\mathbb{R}^n$, in global form, in terms of the standard spatial coordinates in $\mathbb{R}^n$. Such BVPs arise in a variety of situations and have many practical applications. See, for example, \cite[\S{72}]{Ha1} for the heat conduction by surfaces, \cite[\S{10}]{Ar1} for the equations of surface flow, \cite{Ci1}, {\cite{AC1} for the vacuum Einstein equations describing gravitational fields}, \cite{TZ1} for the Navier-Stokes equations on spherical domains, as well as the references therein.

A hypersurface $\mathcal{C}$ in $\mathbb{R}^3$ has the natural structure of a $2$-dimensional Riemannian manifold and the aforementioned PDE's are not the immediate analogues of the ones corresponding to the flat, Euclidean case, since they have to take into consideration geometric characteristics of $\mathcal{C}$ such as curvature. Inherently, these PDE's are originally written in local coordinates, intrinsic to the manifold structure of $\mathcal{C}$.

The operator $\mathcal{D}$ is globally defined on $\mathcal{C}$, and has a relatively simple structure. In terms of (\ref{eq1.1}), the Laplace-Beltrami operator on $\mathcal{C}$ simply becomes {(see \cite[pp. 2ff and p. 8.]{MM1})}
\begin{equation}\label{eq1.2}
\Delta_{\mathcal{C}}=\mathcal{D}^*\mathcal{D}
\quad\mbox{ on }\quad\mathcal{C}.
\end{equation}

\noindent Alternatively, this is the natural operator associated with the Euler-Lagrange equations for the variational integral

\begin{equation}\label{eq1.3}
\mathcal{E}[u]=-\frac12\int_{\mathcal{C}}\|{\mathcal{D}}u\|^2\,dS.
\end{equation}

A similar approach, based on the principle that, at equilibrium,
the displacement minimizes the potential energy, leads to the derivation
of the equation for the elastic hypersurface (cf. \cite{DMM1,Du3} for the
isotropic case).

These results are useful in numerical and engineering applications (cf. \cite{AN1}, \cite{Be1}, \cite{Ce1}, \cite{Co1}, \cite{DL1}, {\cite{BGS1}, \cite{Sm1}}) and we plan to treat a number of special surfaces in greater detail in a subsequent publication.

We consider heat conduction by an "isotropic" media, governed by the Laplace equations and with the classical Dirichlet-Neumann mixed boundary conditions on the boundary in the layer domain $\Omega^\ve:=\cC \times(-\ve,\ve)$ of thickness $2\ve$:
Let us consider the mixed BVP with zero Dirichlet but non-zero Neumann data:
\begin{eqnarray}\label{e7.37}
\begin{array}{lll}
&\Delta_{\Omega^\ve} T(\cx,t)=f(\cx,t), &(\cx,t)\in\cC\times(-\ve,\ve),\\[2mm]
& T^+(\cx,t)=0, &(\cx,t)\in \pa\cC\times(-\ve,\ve),\\[2mm]
&\pm(\pa_t T)^+(\cx,\pm\ve)=q(\cx,\pm\ve),\qquad &\cx\in\cC,
\end{array}
\end{eqnarray}
where $\pm\pa_t=\pa_\nub$ represents the normal derivative on the surfaces $\cC\times{\pm\ve}$. Here $\cC\subset\cS$ is a smooth subsurface of a closed hypersurface $\cS$ with smooth nonempty boundary $\pa\cC$. In the investigation we apply that the Laplace operator
$\Delta_{\Omega^\ve}=\pa^2_1+\pa^2_2+\pa^2_3$ is represented as the sum of the Laplace-Beltrami operator on the mid-surface and the square of the transversal derivative:
$$
\Delta_{\Omega^\ve}T=\sum\limits_{j=1}^{4}\cD_j^2 T=\Delta_{\cC}T+\pa_t^2 T.
$$

The BVP \eqref{e7.37} can be reformulated as the minimization problem for the functional
\begin{eqnarray}\label{e7.38}
E(T_\ve):&\hskip-3mm=&\hskip-3mm\int\limits_{-1}^1\int\limits_{\cC}\Big[\frac12
     \left[\left|(\cD_{\cC}T)(\cx,\tau)\right|^2+\left|(\pa_\tau T)(\cx,\tau)\right|^2\right]+f(\cx,\tau)T(\cx,\tau)\Big]d\sigma d\tau\nonumber\\
&&+\int\limits_{\cC}\left[q(\cx,+\ve)T^+(\cx,+\ve)-q(\cx,-\ve)T^+(\cx,-\ve)\right]d\sigma.
 \end{eqnarray}

After scaling (stretching the variable $t=\ve \tau$ and dividing the entire functional by $\ve$) has the following form
\begin{eqnarray}\label{e7.38a}
E_{\ve}(T_\ve):&\hskip-3mm=&\hskip-3mm\int\limits_{-1}^1\int\limits_{\cC}\Big[\frac12
     \left[\left|(\cD_{\cC}T_\ve)(\cx,\tau)\right|^2+\frac1{\ve^2}\left|(\pa_\tau T_\ve)(\cx,\tau)\right|^2\right]+f_\ve(\cx,\tau)T_\ve(\cx,\tau)\Big]d\sigma d\tau\nonumber\\
&&+\frac1\ve\int\limits_{\cC}\left[q^+(\cx,+\ve)T^+_\ve(\cx,+1)
     -q(\cx,-\ve)T^+_\ve(\cx,-1)\right]d\sigma,\\
&&T_\ve(\cx,\tau):=T\left(\cx,\ve\tau\right)\in \mathbb{H}^1(\Omega^1), \qquad f_\ve(\cx,\tau):=f\left(\cx,\ve\tau\right)\in \mathbb{H}^{-1}(\Omega^1).\nonumber
 \end{eqnarray}

The main result of the present investigation is the following Theorem \ref{t0.1}.
 %
\begin{theorem}\label{t0.1}
Let $q(\cx,\pm\ve)\in\bH^{1/2}(\cC)$ and are uniformly bounded in $\bL_2(\cC)$,
$f_{\varepsilon}(\cx,t)\to f^0(\cx)$ in $\mathbb{H}^{-1}(\Omega^1)$ and there exists a function $q^0\in\bH^{-1/2}(\cC)$ such that
\begin{eqnarray}\label{e7.39}
\lim_{\ve\to0}\frac1{2\ve}\scal\varphi(\cdot),q(\cdot,\ve)-q(\cdot,-\ve)\scar_\cC
     =\scal\varphi,q^0\scar_\cC,\qquad \forall\,\vf\in\bH^{1/2}(\cC).
\end{eqnarray}
Then the functional in \eqref{e7.38a} $\Gamma$-converges to the functional
\begin{eqnarray}\label{e7.40}
E^{(0)}(T)=\int\limits_{-1}^1\int\limits_{\cC}
     \Big[\frac12\left|(\cD_{\cC}T)(\cx,0)\right|^2+[f^0(\cx)+q^0(\cx)]
     T(\cx,0)\Big]d\sigma dt\nonumber\\
=2\int\limits_{\cC}\left[\frac12\left\langle\cD_{\cC} T(\cx),\cD_{\cC}
     T(\cx)\right\rangle + [f^0(\cx)+q^0(\cx)]T(\cx)\right]d\sigma.
\end{eqnarray}

The following Dirichlet boundary value problem for Laplace-Beltrami equation on the mid surface $\cC$
\begin{eqnarray}\label{e7.42}
\begin{array}{l}
\Delta_\cC T(\cx)=f^0(\cx)+q^0(\cx) \quad \cx \in \cC,\\[3mm]
T^+(\cx)=0, \qquad \cx \in \pa\cC.
\end{array}
 \end{eqnarray}
is an equivalent reformulation of the minimization problem with the energy functional \eqref{e7.40} and, therefore, can be considered as the limit of the initial BVP \eqref{e7.37}.
\end{theorem}
 %
\begin{remark}\label{r0.2}
It is remarkable to note that the weak derivative $q^0$ of the Neumann condition from the initial BVP \eqref{e7.37} (see the comment below) migrated into the right hand side of the limit equation.

Let us comment on the condition \eqref{e7.39}. We remind that $q(\cdot,\pm\ve)\in \bH^{1/2}(\cC)$. If $q_1\in\bH^1(\cC \times(-\ve,\ve))$ is an extension of these functions inside the domain $\lim\limits_{t\to\pm\ve}q_1(\cx,t) =q(\cx,\pm\ve)$, then $q^0(\cx)=\dst\frac12(\pa_tq)(\cx,0)\in\bL_2(\cC)$ represents the derivative in the weak sense
 \[
\lim_{\ve\to0}\frac1{2\ve}\scal\vf(\cdot),q(\cdot,\ve)-q(\cdot,-\ve)\scar_\cC
     =\lim_{\ve\to0}\frac1{2\ve}\int_{-\ve}^\ve\scal\vf(\cdot),\pa_\tau q_1)(\cdot,\tau)\scar_\cC\,d\tau=\scal\vf,q^0\scar_\cC
 \]
for all $\vf\in\bL_2(\cC)$ (see Corollary \ref{c7.8} of the Lebesgue Differentiation Theorem below).
 \end{remark}

The layout of the paper is as follows. In \S\,1-\S\,2 we review some basic differential-geometric concepts which are relevant for the work at hand (e.g., hypersurfaces and different methods of their identification). In \S\,3 we identify the most important partial differential operators on hypersurfaces, such as gradient, divergence, Laplace-Beltrami operator. In \S\,4 we consider the energy functional \eqref{eq1.3} and the associated Euler-Lagrange equation. In sections \S\,5, \S\,6
we apply the aforementioned approach and prove main theorems of the present paper, including Theorem \ref{t0.1}.

 %
 %
\section{Brief review of the classical theory of hypersurfaces}
 \label{sec2}
\setcounter{equation}{0}

We commence with the definition of a hypersurface and give two equivalent definitions. Both definitions are important for our purposes.
 %
 \begin{definition}\label{d2.3}
A Subset ${\mathcal S}\subset\mathbb{R}^n$ of the Euclidean space
is called a {\bf hypersurface} if it has a covering ${\mathcal
S}=\bigcup_{j=1}^M{\mathcal S}_j$ and coordinate mappings
 \begin{equation}\label{e1.1}
\Theta_j\;:\;\omega_j\rightarrow{\mathcal S}_j:=\Theta_j(\omega_j)
     \subset\mathbb{R}^n, \qquad\omega_j\subset\mathbb{R}^{n-1},
     \quad j=1,\ldots,M,
 \end{equation}
such that the corresponding differentials
 \begin{equation}\label{eA2.1.15}
\hskip-1.8mm\begin{array}{c} D\Theta_j(p)
      :=\matr\,[\pa_1\Theta_j(p),\ldots,\pa_{n-1}\Theta_j(p)]\, ,
 \end{array}
\end{equation}
have the full rank
 \[
{\rm rank}\,D\Theta_j(p)=n-1\, ,\qquad \forall p\in Y_j\, ,\quad
     k=1,\ldots,n\, ,\quad j=1,\ldots,M\, ,
 \]
i.e. , all points of $\omega_j$ are regular for $\Theta_j$ for all
$j=1,\ldots,M$.

Such mapping is called an {\bf immersion} as well.
 \end{definition}

The hypersurface is called {\bf smooth} if the corresponding
coordinate diffeomorphisms $\Theta_j$ in \eqref{e1.1} are smooth
($ C^\infty$-smooth). Similarly is defined a {\bf $\mu$-smooth}
hypersurface.

The derivatives
\begin{equation}
{\boldsymbol{g}}_{k}({\scriptstyle{\mathcal{X}}})=\partial _{k}\Theta
_{j}(\Theta _{j}^{-1}({\scriptstyle{\mathcal{X}}})),\qquad {\scriptstyle{%
\mathcal{X}}}\in \mathcal{C},\quad k=1,\ldots ,n-1  \label{e1.84}
\end{equation}%
are then tangential vector fields on ${\mathcal{C}}$ and moreover, compose a basis in the space of tangential vector fields $\mathcal{W}({\mathcal{C}})$.

The most important role in the calculus of tangential differential operators we are going to apply belongs to the unit normal vector field $\nub(y)$, $t\in\mathcal{C}$. The {\bf unit normal vector field} to the surface ${\mathcal{C}}$, also known as the {\bf Gau\ss \ mapping}, is defined by the vector product of the covariant basis
\begin{equation} \label{e1.88}
\nub(\cx):=\pm \frac{\boldsymbol{g}_1(\cx)\wedge\ldots\wedge\boldsymbol{g}_{n-1}(\cx)}{
|\boldsymbol{g}_1(\cx)\wedge\ldots\wedge\boldsymbol{g}_{n-1}(\cx)|},\qquad\cx\in\mathcal{C}.
\end{equation}

The system of tangential vectors $\left\{ {\boldsymbol{g}}_{k}\right\}_{k=1}^{n-1}$ to ${\mathcal{C}}$ (cf. \eqref{e1.84}) is, by the definition, linearly independent and is known as the \textbf{covariant basis}. There exists the unique system $\left\{ {\boldsymbol{g}}^{k}\right\} _{k=1}^{n-1}$ biorthogonal to it-the \textbf{contravariant basis}:
\begin{equation*}
\langle {\boldsymbol{g}}_{j},{\boldsymbol{g}}^{k}\rangle =\delta _{jk}\qquad
j,k=1,\ldots ,n-1.
\end{equation*}
The contravariant basis is defined by the formula:
\begin{equation} \label{e2.5}
{\boldsymbol{g}}^k=\frac{1}{\det \,G_\mathcal{S}}{\boldsymbol{g}}_1\wedge
\cdots \wedge{\boldsymbol{g}}_{k-1}\wedge{\boldsymbol{\nu}} \wedge{%
\boldsymbol{g}}_{k+1}\wedge \cdots \wedge{\boldsymbol{g}}_{n-1},\quad
k=1,\ldots ,n-1,
\end{equation}
where
\begin{equation*}
G_{\mathcal{S}}({\scriptstyle{\mathcal{X}}}):=[\langle {\boldsymbol{g}}_{k}({%
\scriptstyle{\mathcal{X}}}),{\boldsymbol{g}}_{m}({\scriptstyle{\mathcal{X}}}%
)\rangle ]_{n-1\times n-1},\qquad p\in \mathcal{S}
\end{equation*}%
is the {\bf Gram matrix}.

Next we expose yet another definition of a hypersurface-an {\bf implicit} one.
 %
 \begin{definition}\label{d2.2}
Let $k\geq1$ and $\omega\subset\bR^n$ be a compact domain. An implicit
$ C^k$-smooth (an implicit Lipschitz) hypersurface in $\bR^n$ is
defined as the set
 \begin{equation}\label{e2.2}
\cS=\Big\{\cx\in \omega\;:\; \Psi_\cS(\cx)=0\Big\}\, ,
 \end{equation}
where $\Psi_\cS\,:\,\omega\rightarrow \bR$ is a $ C^k$-mapping (or is a
Lipschitz mapping) which is regular $\nabla\,\Psi(\cx)\not=0$.
 \end{definition}

\setlength{\unitlength}{0.4mm}
\vskip15mm
\hskip20mm
\begin{picture}(300,140)
\put(-00,40){\epsfig{file=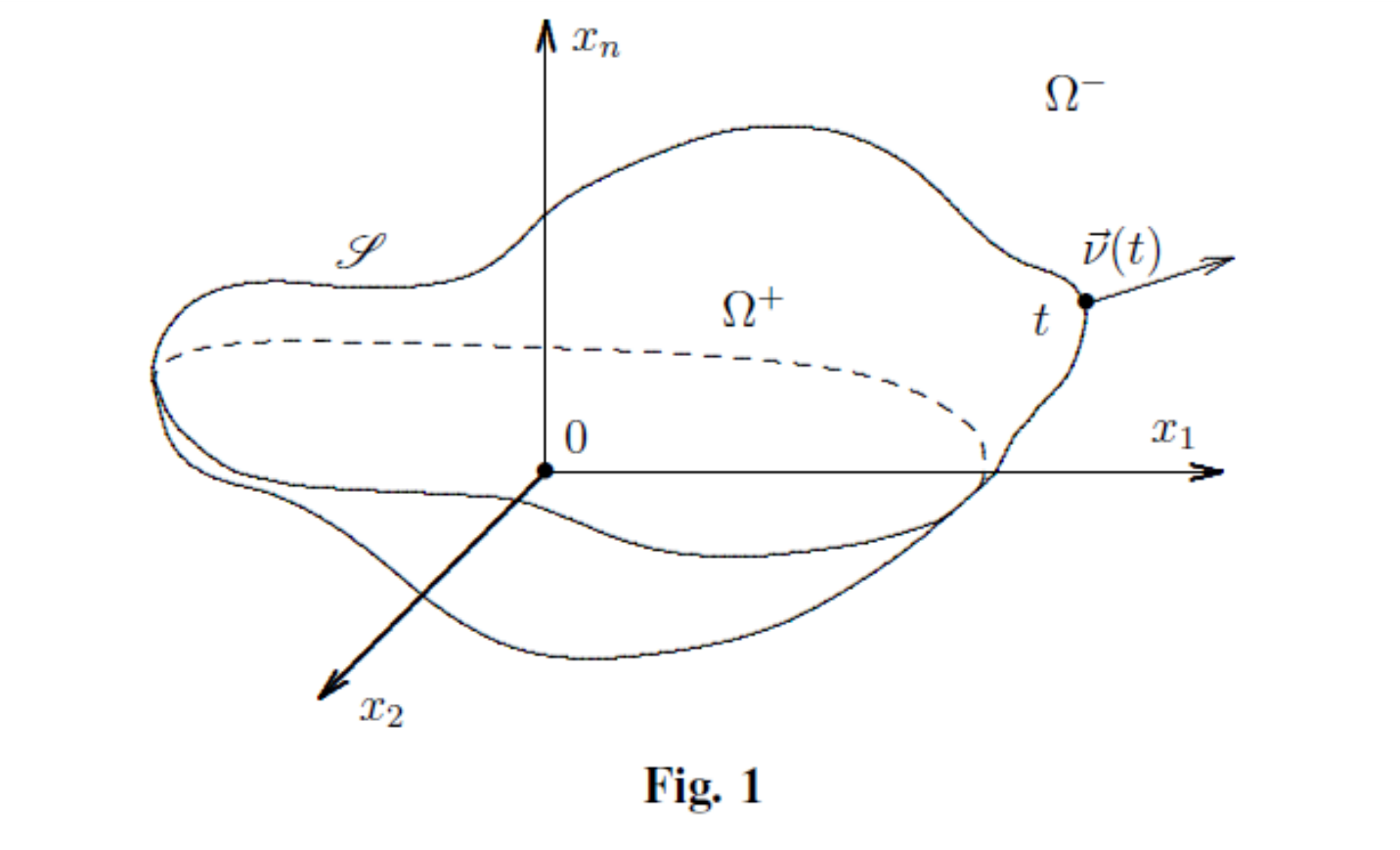, height=60mm, width=120mm}}
\end{picture}
\vskip-15mm

Note, that Definition \ref{d2.3} and Definition \ref{d2.2}  of a hypersurface $\cS$ are equivalent and by taking a single function $\Psi_\cS$ for the implicit definition of a hypersurface $\cS$ we does not restrict the
generality (see e.g., \cite{Du4}).

It is well known that using implicit surface functions gradient (see \eqref{e2.2}) we can write an alternative definition of the unit normal vector field on the  surface (see \eqref{e1.88}):
\begin{equation}\label{e2.4}
\nub(y):=\lim_{x\to t}\frac{(\nabla\Psi _{\mathcal{S}})({x})}{|(\nabla
     \Psi_{\mathcal{S}})(x)|},\qquad t\in\mathcal{S}.
\end{equation}

In applications it is necessary to extend the vector field $\nub(t)$ in a neighborhood of $\cS$, preserving some important features. Here is the precise definition of extension.
 %
 \begin{definition}\label{d2.4.i}
Let ${\cS}$ be a surface in $\bR^n$ with unit normal $\nub$. A vector filed $\cN\in  C^1(\Omega^\ve)$ in a neighborhood $\Omega^\ve$ of ${\cS}$, will be referred to as a {\bf proper extension} if ${\cN}\Big|_\cS={\nub}$, if it is unitary $|{\cN}|=1$ in $\Omega^\ve$ and ${\cN}$ satisfies the following condition in the neighborhood
 \begin{equation}\label{e2.2.19x}
\pa_j\cN_k(x)=\pa_k\cN_j(x)\qquad \mbox{for all}\quad x\in\Omega^\ve,\quad j,k=1,\ldots,n.
 \end{equation}
 \end{definition}

Such extension is needed, for example, to define correctly the normal derivative (the derivative along normal vector fields, outer or inner). It turned out that the "naive" extension (cf. \eqref{e2.4})
\begin{equation}\label{e2.4x}
\nub(t):=\frac{(\nabla\Psi _{\mathcal{S}})({x})}{|(\nabla
     \Psi_{\mathcal{S}})(x)|},\qquad x\in\Omega^\ve
\end{equation}
is not proper (see \cite{DST1} for a counterexample).

For the proof of the next Proposition \ref{p2.4} and Corollary \ref{c2.5} on extension of the normal vector field we refer to \cite{DST1}.
 %
\begin{proposition}\label{p2.4}
Let ${\mathcal{S}}\subset \mathbb{R}^{n}$ be a hypersurface given by an implicit
function
 \[
\mathcal{S}=\left\{\cx\in\mathbb{R}^n\;:\;\Phi_\mathcal{S}(\cx)=0\right\}
 \]
for some  $\Phi_\mathcal{S}\in C^1(\Omega^\ve)$. Then the gradient $\nabla\Phi_\mathcal{S}(x)$ of the function
\begin{equation}\label{e4.3}
\Phi_\mathcal{S}(\cx+t\nub(\cx)):=t, \qquad\cx+t\nub(\cx)\in\Omega^\ve,
\end{equation}
defined in the parameterized neighborhood
 \[
\Omega^\ve:=\left\{x=\cx+t\nub(\cx)\;:\;\cx\in \mathcal{S},\quad
     -\varepsilon<t<\varepsilon\right\}
 \]
for sufficiently small $\varepsilon$, represents a unique proper extension of the unit normal vector field on the surface
 \[
\nub(\cx)=\lim_{x\to\cx}\nabla\Phi_\mathcal{S}(x),\qquad \cx\in\mathcal{S}.
 \]
\end{proposition}
 %
 \begin{corollary}\label{c2.5}
For any proper extension $\cN(x)$, $x\in\Omega^\ve\subset\bR^n$ of the unit normal vector field $\nub$ to the surface $\cS\subset\Omega^\ve$ the equality
 \begin{equation}\label{eq:N}
\pa_\cN\cN(x)=0 \qquad\mbox{holds \ for \ all}\quad x\in\Omega^\ve.
 \end{equation}

In particular, for the derivatives
 \begin{equation}\label{e2.7}
\cD_k=\pa_k-\cN_k\pa_\cN, \qquad k=1,\ldots,n\, ,
 \end{equation}
which are extension into the domain $\Omega^\ve$ of G\"unter's
derivatives $\cD_k=\pa_k-\nu_k\pa_{\nub}$ on the surface $\cS$,
we have the equality:
 \begin{eqnarray}\label{eq:D}
\cD_k\cN_j=\pa_k\cN_j-\cN_k\pa_\cN=\pa_k\cN_j, \qquad \cD_k\cN_j=\cD_j\cN_k,\\
 \text{for all}\quad j,k=1,\ldots,n.\nonumber
 \end{eqnarray}
 \end{corollary}

In the sequel we will dwell on a proper extension and apply the above properties of $\cN$.

Important role in surface geometry goes to the {\bf Weingarten matrix}
 \begin{eqnarray}\label{e2.12}
\cW_\cS(\cx):=\left[\cD_j\nu_k(\cx)\right]_{n\times n},\qquad \cx\in\cS,
 \end{eqnarray}
which is, due to the second equality in \eqref{eq:D}, a symmetric matrix. The mean trace of the Weingarten matrix is a {\bf mean curvature of the surface}:
 \begin{eqnarray}\label{e2.13}
\begin{array}{l}
\cH(\cx):=\dst\frac1{n-1}\text{Tr}\,\cW_\cS(\cx)=\dst\frac1{n-1}
     \dst\sum\limits_{k=1}^n\cD_k\nu_k(\cx)=\dst\sum\limits_{k=1}^n
     \lambda_k(\cx),\qquad \cx\in\cS,
\end{array}
 \end{eqnarray}
where $\lambda_1(\cx),\ldots,\lambda_n(\cx)$ are the eigenvalues of $\cW_\cS(\cx)$. The Weingarten matrix is degenerated
 \[
\det\,\cW_\cS(\cx)\equiv0\qquad \text{for all}\quad \cx\in\cS
 \]
because $\cW_\cS(\cx)\nub(\cx)\equiv0$. Therefore one of the eigenvalues is zero, say $\lambda_n(\cx)\equiv0$ for all $\cx\in\cS$. The {\bf Gauss curvature} of the surface coincides with the product of non-degenerated eigenvalues of the Weingarten matrix:
 \begin{eqnarray}\label{e2.14}
\cG_\cS(\cx):=\lambda_1(\cx)\cdots\lambda_{n-1}(\cx),\qquad \cx\in\cS
 \end{eqnarray}
(cf. \cite{Du4,Du5,DK1} for details).

 %
 %
 \section{Calculus of tangential differential operators}
 \label{sec3}

The content of the present section partly follows \cite[\S\, 4]{DMM1} and \cite[\S\S\, 4,5]{Du4}.

In the present section we consider a hypersurface $\cS$, which is the boundary of some domain $\Omega\subset\bR^n$. The boundary hypersurface $\cS$ is given by an immersion \eqref{e1.1}. For the sake of simplicity we drop the indices $\Theta_1,\ldots,\Theta_m$, $\omega_1,\ldots,\omega_m$, but will resume indexing if necessary.  $\nub(t)=(\nu_1(t),\ldots, \nu_n(t))^\top$ is the outer (with respect to $\Omega$) unit normal vector field to $\cS$
(cf. \eqref{e2.4} and \eqref{e1.88}) and $\cN(x)$ is the proper extention of $\nub$ in a neighborhood $\omega_\cS$ of $\cS$ (cf. Definition \ref{d2.4.i}).

A hypersurface $\cC$ is a part of $\cS$ and has a smooth boundary $\Gamma=\pa\cC$, given by another immersion
 \begin{eqnarray}\label{e2.5ay}
\Theta_\Gamma\,:\,\partial\omega\rightarrow\Gamma:=\pa\cC\, ,
     \qquad\partial\omega\subset\bR^{n-2}.
 \end{eqnarray}
$\nub_\Gamma(t)$ is the outer normal vector field to the boundary $\Gamma$, which is tangential to $\cC$ (and to $\cS$).

By $\cV(\cS)$ we denote the set of all smooth vector fields, tangential to
the hypersurface $\cS$:
 \begin{equation}\label{eA1.1.1}
\U\;:\;\omega\rightarrow\bR^n\, ,\qquad \U(x)=\sum\limits_{j=1}^n
     U_j(x)\e^j\, ,\qquad \langle\U(\cx),\nub(\cx)\rangle\equiv0
 \end{equation}
where $U^j\in C^\infty_0(\cS)$ and $\{\e^j\}_{j=1}^n$ is the natural Cartesian basis in $\bR^n$
 \begin{equation}\label{e1.1.x}
\e^1:=(1,0,\ldots,0),\ldots, \e^n:=(0,\ldots,0,1),
 \end{equation}
while $\langle x,y\rangle$ denotes the scalar product in $\mathbb{R}^n$:
 \[
\langle x,y \rangle:=\sum_{j=1}^nx_jy_j,\qquad x,y\in\mathbb{R}^n.
 \]

A {\bf curve} on a smooth surface $\cS$ is a mapping
 \begin{equation}\label{eA2.1.1}
\gm\;:\;\cI\mapsto\cS\, ,\qquad \cI:=[0,1]\subset\bR\, ,
 \end{equation}
of a line interval $\cI$ to $\cS$.

Let $\U\in\cV(\cS)$ and consider the corresponding ordinary differential equations (ODE):
 \begin{equation}\label{eA1.1.3}
y'=\U(y)\, ,\qquad y(0)=\cx\, ,\qquad \cx\in\cS\, .
 \end{equation}
A solution $y(t)$ of \eqref{eA1.1.3} is called an {\bf integral curve}
(or {\bf orbit}) of the vector field $\U$ and represents a subset of the surface $\cS$. The mapping
 \begin{equation}\label{eA1.1.4}
y=y(t,\cx)=\cF^t_{\U}(\cx)\;:\;\cI\times\cS\rightarrow\cS\subset\bR^n,\qquad
     \cI:=[0,1],
 \end{equation}
is called the {\bf flow} generated by the vector field $\U$ at the point $\cx$.

A vector field $\U\in\cV(\Omega)$ defines the {\bf first order differential operator}
 \begin{equation}\label{eA1.1.5}
\U f(\cx)=\partial_{\U} f(\cx):=\lim\limits_{h\to0}\frac{f\left(
     \cF^h_{\U}(\cx)\right)-f(\cx)}{h}=\frac d{dt}\left.f\left(
     \cF^t_{\U}(\cx)\right)\right|_{t=0}
 \end{equation}
for a function defined on the surface $\cS$, which is called the derivative along $\U$. If $f(x)$ is defined in the neighbourhood of the surface $\cS$, by applying the chain rule to \eqref{eA1.1.5} we get
 \begin{eqnarray}\label{eA1.1.7}
\partial_{\U} f(x)=\langle \U(x),\nabla f(x)\rangle
     =\sum\limits_{j=1}^nU_j(x)\frac{\pa f}{\pa x_j}.
 \end{eqnarray}

In particular, the G\"unter's derivatives
 \[
\cD_j:=\pa_j-\nu_j\partial_{\nub}=\pa_j-\nu_j\sum_{k=1}^n\nu_k\pa_k\, ,
     \qquad j=1,\ldots,n,
 \]
introduced in \eqref{e2.7}, are tangential. Another set of tangential derivatives on the surface $\cS$ is the Stokes' derivatives
 \begin{equation}\label{e3.14}
\cM_{jk}=\nu_j\pa_k-\nu_k\pa_j,\qquad j,k=1,\ldots,n.
 \end{equation}
Gunter's and Stockes derivatives differentiate functions along the following tangent vector fields
 \begin{eqnarray}\label{e2.x3}
 \begin{array}{c}
\cD_j:=\partial_{\d^j}=\d^j\cdot\nabla\, ,\qquad
     \cM_{jk}:=\partial_{\fm_{jk}}=\fm_{jk}\cdot\nabla\, ,\\[2mm]
\d^j:=\pi_\cS \e^j=\e^j-\nu_j\nub,\quad \fm_{jk}:=\nu_j\e_k-\nu_k\e_j\, ,\\[2mm]
     \langle \d^j,\nub\rangle=0\, ,\qquad \langle\fm_{jk},\nub\rangle=0\, ,\quad j,k=1,\ldots,n\, .
 \end{array}
 \end{eqnarray}
The following reciprocal representations are easy to verify:
 \begin{eqnarray}\label{e3.16}
\cD_j:=\sum \nu_k\cM_{kj},\qquad\cM_{jk}=\nu_j\cD_k-\nu_k\cD_j,\qquad j,k=1,\ldots,n.
 \end{eqnarray}

The generating vector fields $\big\{\d^j\big\}_{j=1}^n$ and $\big\{\fm_{jk}\big\}_{j,k=1}^n$ are not bases in the space of tangential vectors to $\cS$, since they are linearly dependent
 \begin{eqnarray}\label{e2.x4}
\sum_{j=1}^n\nu_j(\cx)\d^j(\cx)\equiv0, \qquad \fm_{jj}=0, \qquad \fm_{jk}=-\fm_{kj},
 \end{eqnarray}
but both systems $\big\{\d^j\big\}_{j=1}^n$ and
$\big\{\fm_{jk}\big\}_{0\leqslant j<k\leqslant n}$ are full and any tangential vector field $\U\in\cV(\cS)$ is represented as follows
 \begin{eqnarray}\label{e2.x5}
\U(\cx)=\sum_{j=1}^nU^j(\cx)\d^j(\cx)=\sum_{0\leqslant j<k\leqslant n}^nc_{jk}
     (\cx)\fm_{jk}(\cx)\, .
 \end{eqnarray}

For a properly extended normal vector field  $\cN$ (cf. Definition \ref{d2.4.i}). we can extend the operators $\cD_j$ and $\cM_{jk}$ (cf. \eqref{e2.7})
 \begin{eqnarray}\label{e3.19}
\cD_j=\pa_j-\cN_j\partial_\cN\, , \qquad \cM_{jk} :=\cN_j\pa_k
     -\cN_k\pa_j\, ,\qquad 1\leqslant j,k\leqslant n
 \end{eqnarray}
In the sequel, we shall make no distinction between the operator
$\cD_j$ or $\cM_{jk}$ on $\cS$ and the extended one in $\bR^n$
given by \eqref{e3.19}.

Throughout the paper we use the following notation for the scalar products
 \begin{equation}\label{e4}
\scal u,v\scar_\cS:=\oint\limits_\cS u^\top(t)\ov{v(t)}d\sigma\, ,
    \qquad \scal\vf,\ v\scar_\Gm:=\oint\limits_\Gm\vf^\top(s)\ov{\ v(s)}d\fs\, .
\end{equation}

For a tangential differential operator $P$ on a closed hypersurface $\cS$ let $P_\cS^*$ denote the ``surface'' adjoint:
\begin{eqnarray}\label{e10.6.0}
\scal P\vf,\psi\scar_\cS:=\oint\limits_{\cS}\langle P\vf,
    \psi\rangle\,d\sigma=\oint\limits_{\cS}\langle\vf,P^*_\cS
    \psi\rangle\,d\sigma=\scal\vf,P^*_\cS \psi\scar_\cS\\
    \forall\,\vf,\,\psi\in C^1(\cS). \nonumber
\end{eqnarray}

In \cite{DMM1} is shown that for a tangential differential operator $P\vf=\sum_{j=1}^na_j\pa_j \vf + b\vf$ the surface-adjoint and the formally adjoint operators coincide, i.e.,
 \begin{equation}\label{e2.a}
P^*_\cS\vf=P^*\vf=-\sum_{j=1}^n\pa_ja^\top_j\vf +b^\top\vf\, .
 \end{equation}

In particular, the Stokes' derivatives are skew-symmetric
\begin{eqnarray}\label{e2.b}
\big(\cM^*_{jk}\big)_\cS=\cM^*_{jk}=-\cM_{jk}=\cM_{kj}\qquad
    \forall\,j,k=1,\ldots,n\, ,
\end{eqnarray}
while the adjoint operator to the operator $\cD_j$ is given by formula
 \begin{equation}\label{e2.4.13}
\big(\cD_j\big)^*_\cS\vf=\cD_j^*\vf=-{\cD}_j\vf-\nu_j\cH^0_\cS\vf\, ,
    \qquad \vf\in C^1(\cS),
 \end{equation}
where $\cH^0_\cS(\cx)=(n-1)\cH_\cS(\cx)$ is proportional to the mean curvature (see \eqref{e2.13}).
 %
 \begin{proposition}[\cite{DMM1} Theorem 5.1, \cite{Du4}, Theorem 4.1]\label{t4.1}
The surface gradient and the surface divergence represented in Gunter's derivatives have the following form
 \begin{eqnarray}\label{e4.1}
\nabla_\cS\vf =\Bigl\{{\cD}_1\vf,{\cD}_2\vf,...,
     {\cD}_n\vf\Bigr\}^\top,\\
\label{e4.2}
\Div_{\cS}\,\V=-\nabla^*_\cS \V:=\sum\limits_{j=1}^n\cD_jV^j,
 \end{eqnarray}
where $\vf\in C^1(\cS)$ is a scalar function and $\V=\sum_{j=1}^nV^je_j\in\cV(\cS)$ is a 1-smooth tangential vector field. The Laplace-Beltrami operator $\Dlb_{\cS}$ on $\cS$ has the form
 \begin{eqnarray}\label{e2.5.23}
\Dlb_\cS\,\psi=\sum\limits_{j=1}^n\cD_j^2\psi=\sum_{j<k}\cM_{jk}^2\psi
     =\frac12\sum_{j,k=1}^n\cM_{jk}^2\psi\qquad\forall\,\psi\in C^2(\cS)\, .
 \end{eqnarray}
 \end{proposition}

The following Proposition \ref{p3.2} is important while considering boundary value problems for Laplace-Beltrami equation (cf. \cite{Du3} for a proof).
 %
 \begin{proposition}\label{p3.2}
For $\vf\in C^1(\cS)$ the surface gradient vanishes $\nabla_\cS\vf\equiv0$ if and only if $\vf(\cx)\equiv\const$.
 \end{proposition}

Let $1<p<\infty$, $s\in\bR$. For the definition of  Bessel potential $\bH^s_p(\cS)$ and Sobolev-Slobodeckii $\bW{}^s_p(\cS)$ spaces for a closed smooth manifold $\cS$ we refer to \cite{Tr1} (also see \cite{Du2,Hr1} etc.). For $p=2$ the Sobolev--Slobodetski $\bW^s_2(\cS)$ and Bessel potential $\bH^s_2(\cS$ spaces coincide (i.e., the norms are
equivalent). For an integer $m=1,2,\ldots$ the spaces $\bW{}^m_p(\cS)$ and $\bH^s_p(\cS)$ coincide with the Sobolev space and an equivalent norm in the Sobolev space is defined with the help of Gunter's derivatives (the derivatives are understood in distributional sense)
\[
\|\vf\,\big|\,\bW_p^\ell(\cS)\,\|:=\left[\sum\limits_{|\al|\leq\ell}
    \|\cD_\al\vf\,\big|\,\bL_p(\cS)\|\right]^{1/p}.
\]

By $\bX_p^s(\cS)$ denote one of the following: Bessel potential $\bH^s_p(\cS)$ or Sobolev-Slobodeckii $\bW{}^s_p(\cS)$ space. Consider the space
 \begin{eqnarray}\label{e1.21b}
\bX^s_{p,\#}(\cS):=\left\{\vf\in\bX^s_2(\cS)\;:\;\scal\vf,1\scar_\cS=0\right\}.
 \end{eqnarray}
It is obvious, that $\bX^s_{p,\#}(\cS)$ does not contain constants: if $c_0={\rm const}\in\bX^s_{p,\#}(\cS)$ than
 \[
0=\scal c_0,1\scar_\cS=c_0\scal1,1\scar_\cS=c_0\mes\,\cS
 \]
and $c_0=0$. Moreover, $\bX^s_p(\cS)$ decomposes into the direct sum
 \begin{eqnarray}\label{e1.21c}
\bX^s_p(\cS)=\bX^s_{p,\#}(\cS)+\{{\rm const}\}
 \end{eqnarray}
and the dual (adjoint) space is (see \cite{DTT1} for details)
 \begin{eqnarray}\label{e1.21d}
(\bX^s_{p,\#}(\cS))^*=\bX^{-s}_{p',\#}(\cS), \qquad p':=\frac p{p-1}.
 \end{eqnarray}
 %
 \begin{theorem}\label{t3.3}
Let $\cS$ be an $\ell$-smooth closed hypersurface, $\ell=1,2,\ldots$, $1<p<\infty$ and $|s|\leqslant\ell$. Let $\bX^s_p(\cS)$  be the same as in \eqref{e1.21b}- \eqref{e1.21d}.

Let $\cA$ be a positive definite matrix-function
\begin{equation}\label{e3.26}
\langle \mathcal{A}(\cx)\xi,\xi\rangle\geqslant C\|\xi\|^2, \qquad \xi\in\bR^n
\end{equation}
for all $\cx\in\cC$. Then the "anisotropic" Laplace-Beltrami operator between the spaces with detached constants (see \eqref{e1.21b})
 \begin{eqnarray*}
\Div_\cS(\mathcal{A}\,\nabla_\cS)\;:\;\bX^{s+1}_{p,\#}(\cS)\to\bX^{s-1}_{p,\#}(\cS).
 \end{eqnarray*}
is invertible. Moreover, in the setting
 \begin{eqnarray*}
-\Div_\cS(\mathcal{A}\,\nabla_\cS)\;:\;\bW^1_{2,\#}(\cS)\to\bW^{-1}_{2,\#}(\cS)
 \end{eqnarray*}
the operator is self adjoint and positive definite:
 \begin{eqnarray}\label{e3.28}
&&\hskip-15mm\scal-\Div_\cS(\mathcal{A}\nabla_\cS\vf),\psi\scar_\cS
     =\scal\vf,-\Div_\cS(\mathcal{A}\nabla_\cS\psi)\scar_\cS,\\[2mm]
\label{e3.29}
&&\hskip-15mm\scal-\Div_\cS(\mathcal{A}\nabla_\cS\vf),\vf\scar_\cS\geq
     C\|\vf\big|\bW^1_{2,\#}(\cS)\|^2 \qquad \text{for all} \quad\vf,\psi\in\bW^1_{2,\#}(\cS).
 \end{eqnarray}
 \end{theorem}
{\bf Proof:} For the proof see \cite[Theorem 1.10]{DTT1}).    \QED

Now let $\cC\subset\cS$ be a smooth subsurface of a closed hypersurface
$\cS$ and $\gm=\pa\cC\not=\emptyset$ be its smooth boundary
$\pa\cC=\Gm$ (see Fig. 2).
%
\setlength{\unitlength}{0.4mm}
\vskip25mm
\hskip20mm
\begin{picture}(300,140)
\put(-00,40){\epsfig{file=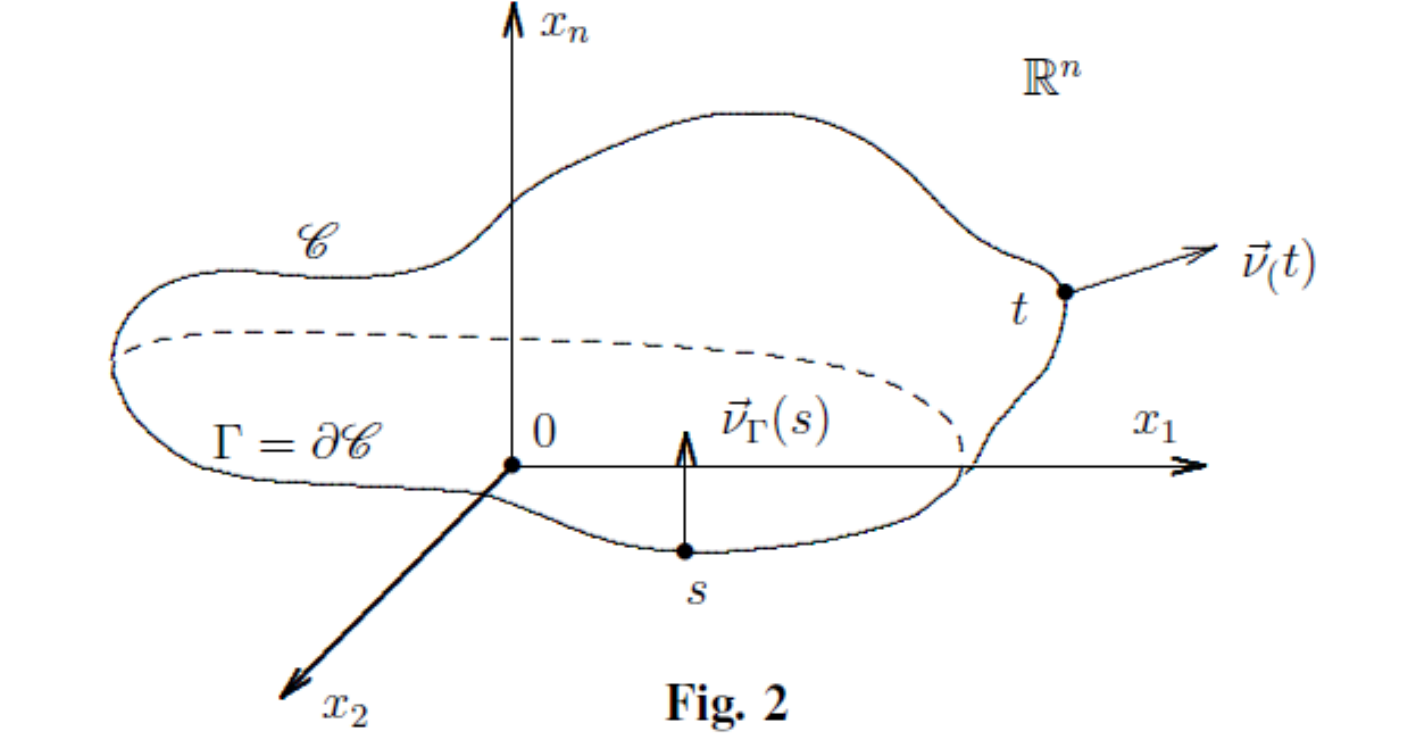, height=60mm, width=120mm}}
\end{picture}
\vskip-15mm

The space $\wt {\bH}_p^s(\cC)$ is defined as a subspace of those functions $\vf\in \bH_p^s(\cS)$, which are supported in the subsurface, $\supp\,\vf\subset\ov{\cC}$, whereas $\bH_p^s(\cC)$ denotes the quotient space $\bH_p^s(\cC)=\bH_p^s(\cS)\Big/\wt{\bH}_p^s(\cC^c)$, where $\cC^c:=\cS\setminus\ov{\cC}$ is the complementary surface to $\cC$. The space $\bH_p^s(\cC)$ can be identified with the space of distributions $\vf$ on $\cC$ which have an extension to a distribution $\ell\vf\in\bH_p^s(\cS)$. Therefore $r_\cC\bH_p^s(\cS)=\bH_p^s(\cC)$, where $r_\cC$ denotes the restriction operator of functions (distributions) from the surface $\cS$ to the subsurface $\cC$.

The spaces $\wt {\bW}_p^s(\cC)$ and $\bW_p^s(\cC)$ are defined similarly (see \cite{Tr1} and also \cite{Du2,Hr1} etc.).

The subspaces $\wt{\bH}_{p,\#}^s(\cC)$, $\wt{\bW}_{p,\#}^s(\cC)$, $\bH_{p,\#}^s(\cC)$ and $ \bW_{p,\#}^s(\cC)$ are defined similarly as in \eqref{e1.21b}: they consist of functions from the corresponding spaces which have mean value zero $\scal\vf,1\scar_\cC=0$.

Let us consider the following boundary value problems for the "anisotropic" Laplace equation with mixed boundary conditions
\begin{equation}\label{e3.32}
\left\{\begin{array}{ll}
\Div_\cC(\mathcal{A}\,\nabla_\cC u)(t)=f(t),\qquad & t\in\cC, \\[0.2cm]
u^+(s)=g(s),     \qquad & {\rm on} \quad  \Gm_D, \\[0.2cm]
\langle\nub_\Gm(s),(\mathcal{A}\,\nabla_\cC u)^+(s)\rangle=h(s),\qquad &{\rm on}\quad \Gm_N,
\end{array}\right.
\end{equation}
where $\pa\cC=\Gm=\Gm_D\cup\Gm_N$ is a decomposition of the boundary in two connected parts and $\mathcal{A}=\{a_{ij}\}$ is $n\times n$ strictly positive definite matrix-function (see \eqref{e3.26}.

The BVP \eqref{e3.32} we consider in the following weak classical setting
\begin{equation}\label{e3.33}
f\in\wt\bH^{-1}(\cC),\qquad g\in\bH^{1/2}(\Gamma_D),\qquad
     h\in\bH^{-1/2}(\Gamma_N).
\end{equation}
 %
\begin{theorem}\label{t3.4}
The mixed boundary value problem \eqref{e3.32} in the weak classical setting
\eqref{e3.33} has a unique solution in the space $\bW^1(\cC)$.
\end{theorem}
{\bf Proof:} For the proof we quote \cite[Theorem 2.2]{DTT1}.    \QED

In conclusion of the present section let us recall the definition of surface $\delta$-function (see, e.g., \cite[(4.30) in \S\, 4]{Du2}).

Let $k=1,2,\ldots$ and $\cC$ be a $C^k$-smooth hypersurface in $\bR^n$, open or closed. The surface $\dl$-function is defined by the equality
\begin{equation}\label{e3.32x}
(g\otimes\delta_\cC,v)_{\cC}:=(g,\gm_\cS v)_{\cC}=\int\limits_\cC g(\tau)\gm_\cS
     v(\tau)d\sigma,\qquad g\in C^k(\cC),\quad v\in C_0^k(\bR^n),
\end{equation}
where $\gm_\cS v(\tau)$ denotes the trace on the boundary surface. Obviously, $\supp(g\otimes\delta_\cC)=\supp\,g\subset\cC$.

In the next Lemma \ref{l3.5} the definition \eqref{e3.32x} is extended to less regular functions.
\begin{lemma}\label{l3.5}
Let $1<p<\infty$ $(1\leq q\leq\infty),$ $s<0$, $g\in{\bH}^s_p(\cC)$ (or $g\in {\bB}^s_{p,q}(\cC)$). Then
\[
g\otimes\delta_\cC\in\bH_p^{s-\frac1{p'}}(\bR^n\setminus\cC)\, , \quad
     \left(g\otimes\delta_\cS\in\bB_{p,q}^{s-\frac1{p'}}(\bR^n\setminus\cC)\right),
\]
where $p'=p/(p-1)$. In particular, if $g\in{\bH}^{-1/2}(\cC)$, than $g\otimes\delta_\cC\in\bH^{-1}(\bR^n\setminus\cC)$.
\end{lemma}

We conclude the section with some auxiliary results on Lebesgue points of integrable functions, which we apply in proofs later in \S\, \ref{sec7}.

Let $B(x)$ be a ball in the Euclidean space $B\subset\mathbb{R}^n$ centered at $x$. {\em The derivative of the integral at $x$} is defined to be
\begin{eqnarray}\label{e3.32a}
\lim_{B(x) \rightarrow x} \frac{1}{|B(x)|} \int_{B(x)}f(y) \, \mathrm{d}y,
\end{eqnarray}
where $|B(x)|$ denotes the volume (i.e., the Lebesgue measure) of $B(x)$, and $B(x)\rightarrow x$ means that the diameter of $B(x)$ tends to $0$. Note that
\begin{eqnarray}\label{e3.32b}
\left|\frac{1}{|B(x)|} \int_{B(x)}f(y)\,dy - f(x)\right| = \left|\frac{1}{|B(x)|} \int_{B(x)}[f(y) -f(x)]\,dy\right|\nonumber\\
\leqslant \frac{1}{|B(x)|} \int_{B(x)}|f(y) -f(x)|\,dy.
\end{eqnarray}
The points $x$ for which the right hand side tends to zero are called the {\em Lebesgue points of $f$.}
 %
\begin{theorem}[Lebesgue Differentiation Theorem, Lebesgue 1910.]\label{t7.7}
For an integrable function $f\in\bL_1(\Omega)$ the derivative of the integral \eqref{e3.32a} exists and is equal to $f(x)$ at almost every point $x\in\Omega$.

Moreover, almost every point $x\in\Omega$ is a Lebesgue point of $f$ (see \eqref{e3.32b}).
\end{theorem}
 %
\begin{corollary}\label{c7.8}
If $g\in\bL_2(\Omega)$, $f\in\bL_2(\Omega\times(-1,1))$, then
\begin{eqnarray}\label{e7.45}
\lim_{\ve\rightarrow0}\frac{1}{2\ve}\int_{t-\ve}^{t+\ve}\scal g(\cdot),f(\cdot,\tau))\scar_{\Omega}d\tau=\scal g(\cdot),f(\cdot,t))\scar_{\Omega}
 \end{eqnarray}
for almost all $t\in(-1,1)$.
\end{corollary}
{\bf Proof:} It is clear, that $g\cdot f\in\bL_1(\Omega\times(-1,1))$ and for the function $h(t):=\scal g(\cdot),f(\cdot,t) \scar_{\Omega}$ the inclusion $h\in\bL_1((-1,1))$ is true. Thence we can apply Theorem \ref{t7.7} to the function $h(t)$ and get \eqref{e7.45}.  \QED

 %
 %
\section{Laplace operator in a layer domain}
 \label{sec4}
\setcounter{equation}{0}

We will keep the notation of \S\,2: $\Theta$, $\Omega^\ve$, $\omega$, $\cS$ and $\cC$. We consider a {\bf layer domain}
 \begin{eqnarray}\label{e5.1}
\Omega^\ve:=\Big\{\cx_t\in\bR^n\,:\,\cx_t=\cx+t\nub(\cx)=\Theta(x)
     +t\nub\big(\Theta(x)\big)\, ,\quad x\in\omega\, ,\;-\ve<t<\ve\Big\}\nonumber\\[2mm]
=\mathcal{C}\times(-\varepsilon,\varepsilon),
 \end{eqnarray}
where $\nub(\cx)=\nub(\Theta(x))$ for $\cx=\Theta(x)\in\cS$, is the
outer unit normal vector field (see \eqref{e1.88} and \eqref{e2.4}). The surface $\cC$ is a mid-surface for the layer domain.

We will also use the notation $\nub(x):=\nub(\Theta(x))$ for brevity unless this does not leads to a confusion. The coordinate $t$ will be referred to as the {\bf transverse variable}.

Without going into detail let us remark only that if the hypersurface $\cS$
is $C^1$-smooth and $1/\ve$ is more than the maximum of modules of all
principal curvatures of the surface $\cS$ (i.e., of all eigenvalues $|\lambda_1(\cx)|,\ldots,|\lambda_{n-1}(\cx)|,|\lambda_n(\cx)|$ of the
Weingarten matrix $\cW_\cS(\cx)$, $\cx\in\cS$), then the mapping
 \begin{eqnarray}\label{e5.2}
 \begin{array}{c}
\Theta^\ve\,:\,\omega^\ve:=\omega\times(-\ve,\ve)\rightarrow \Omega^\ve\, ,
     \quad\omega^\ve\subset\bR^n\, ,\\[2mm]
\Theta^\ve(y,t):=\Theta(y)+t\nub(y)\, ,\qquad (y,t)\in\omega^\ve
 \end{array}
 \end{eqnarray}\index{N}{$\Theta^\ve$}\index{N}{$\omega^\ve$}
is a diffeomorphism.

We will also suppose that $\cN$ is a proper extension of the outer
unit normal vector field $\nub(t)$ into the layer neighborhood $\Omega^\ve$ (cf. Definition \ref{d2.4.i}).

The n-tuple $\g_1:=\pa_1\Theta,\ldots,\g_{n-1}:=\pa_{n-1}\Theta,\g_n:=\cN$, where $\cN$ is the proper extension of $\nub$ in the neighborhood $\Omega^\ve$, is a basis in $\Omega^\ve$ and arbitrary vector field $\U=\sum_{j=1}^nU^0_j\e^j\in\cV(\Omega^\ve)$ is represented with this basis in ``curvilinear coordinates".

Let us consider the system of $(n+1)$-vectors
 \begin{eqnarray}\label{e5.15}
\d\,^j:=\e^j-\cN_j\cN\, ,\qquad j=1,\ldots,n \quad {\rm and}\quad
     \d\,^{n+1}:=\cN,
 \end{eqnarray}
where $\e^1,\ldots,\e^n$ is the Cartesian basis in $\bR^n$ (cf. \eqref{e1.1.x}); the first $n$ vectors $\d\,^1,\ldots,\d\,^n$ are tangential to the surface $\cC$, while the last one $\d\,^{n+1}=\cN$ is orthogonal to all $\d\,^1,\ldots,\d\,^n$. This system is, obviously, linearly dependent, but full and any vector field $\U\in\cV(\Omega^\ve)$ is written in the following form:
 \begin{eqnarray}\label{e5.16}
\U=\dst\sum_{j=1}^nU_j\e^j=\dst\sum_{j=1}^{n+1}U^0_j\d\,^j.
 \end{eqnarray}

Since the system $\big\{\d\,^j\big\}_{j=1}^{n+1}$ is linearly dependent
 \begin{eqnarray}\label{e5.17}
\dst\sum_{j=1}^n\cN_j\d\,^j=0,\qquad \langle\cN_j,\d^j\rangle=0,\quad j=1,\ldots,n,
 \end{eqnarray}
the representation \eqref{e5.16} is not unique. To fix the unique representation in \eqref{e5.16} we will keep the following convention:
 \begin{eqnarray}\label{e5.19}
U^0_j:=U_j-\langle\cN,\U\rangle\cN_j,\quad j=1,\ldots,n,\quad
     U^0_{n+1}=\langle\cN,\U\rangle=\dst\sum_{j=1}^nU_j\cN_j.
 \end{eqnarray}
The convention \eqref{e5.19} is natural because if the  vector $\U(\cx)$ is tangent to $\cC$ for $\cx\in\cC$, than $U^0_j(\cx):=U_j(\cx)$ for $j=1,\ldots,n$ and $U^0_{n+1}(\cx)=0$.

Moreover, if the scalar product of vectors
 \begin{eqnarray}\label{e5.32}
\U:=\dst\sum_{j=1}^nU_j\d\,^j=\dst\sum_{j=1}^{n+1}U^0_j\d\,^j,\quad
          \V:=\dst\sum_{j=1}^nV_j\d\,^j=\dst\sum_{j=1}^{n+1}V^0_j\d\,^j
 \end{eqnarray}
is defined by the equality
 \[
\langle\U^0,\V^0\rangle:=\sum_{j=1}^{n+1}U^0_j\overline{V^0_j},
 \]
than the "new" and the "old" scalar products coincide:
 \begin{eqnarray}\label{e5.33}
\langle\U,\V\rangle^0&\hskip-3mm=&\hskip-3mm\sum_{j=1}^{n+1}U^0_j\overline{V^0_j}
     =\sum_{j=1}^n(U_j-\cN_j\langle\cN,\U\rangle)(\overline{V_j-\cN_j\langle
     \cN,\V\rangle})+\langle\cN,\U\rangle\langle\cN,\V\rangle\nonumber\\
&\hskip-3mm=&\hskip-3mm\sum_{j=1}^nU_j\overline{V_j}=\langle\U,\V\rangle.
 \end{eqnarray}
In particular,
 \begin{equation}\label{e5.34}
\|\U\|^0:=\sum_{j=1}^{n+1}|U^0_j|^2=\sum_{j=1}^n|U_j|^2=\|\U\|.
 \end{equation}

Note for a later use, that due to the equalities \eqref{e5.17} and the convention \eqref{e5.19} we get
 \begin{eqnarray}\label{e5.22}
\pa_{\U}&\hskip-3mm=&\hskip-3mm\dst\sum_{j=1}^nU_j\pa_j=\sum_{j=1}^n[U^0_j\pa_j
     +\langle\cN,\U\rangle\cN_j\pa_j=\sum_{j=1}^nU^0_j(\pa_j-\cN_j\pa_\cN)
     +\langle\cN,\U\rangle\pa_\cN\nonumber\\
&\hskip-3mm=&\hskip-3mm\sum_{j=1}^nU^0_j\cD_j+U_{n+1}\cD_{n+1}
     =\sum_{j=1}^{n+1}U^0_j\cD_j=:\cD_\U.
 \end{eqnarray}
 %
 \begin{definition}\label{d5.6}
For a function $\vf\in\bW^1(\Omega^{\ve})$ the extended gradient is
 \begin{equation}\label{e5.23}
\cD_{\Omega^\ve}\,\vf=\Bigl\{\cD_1\vf,...,\cD_n\vf,
     \cD_{n+1}\vf\Bigr\}^\top
     =\sum_{j=1}^{n+1}(\cD_j\vf)\d^j,\qquad \cD_{n+1}\vf:=\pa_\cN\vf
\end{equation}
and for a smooth vector field $\U=\dst\sum_{j=1}^{n+1}U^0_j\d\,^j\in\cV( \Omega^\ve)$ (see \eqref{e5.16}, \eqref{e5.19}) the extended divergence is
 \begin{eqnarray}\label{e5.24}
\Div_{\Omega^\ve}\,\U:=\dst\sum\limits_{j=1}^{n+1}\cD_jU^0_j
     +\cH^0_\cC\langle\cN,\U\rangle=-\nabla^*_{\Omega^\ve}\U,
 \end{eqnarray}
since
 \begin{eqnarray}\label{e5.25}
\cH^0_{\Omega^\ve}(x):=\sum_{j=1}^n\pa_j\cN_j(x)=\sum_{j=1}^{n+1}\cD_j\cN_j(x)
     =\sum_{j=1}^n\cD_j\nu_j(t)=\cH^0_\cC(t),\\
x\in\Omega^\ve,\qquad t=\pi_\cS x\nonumber
 \end{eqnarray}
and $\cH^0_\cC(t)$ differs from the mean curvature $\cH_{\cC}(t)$ (see \eqref{e2.13}) by the constant multiplier $\cH^0_\cC(t)=(n-1)\cH_\cC(t)$.
 \end{definition}
 %
\begin{lemma}\label{l5.7} The classical gradient $\nabla\vf:=\Bigl\{\pa_1 \vf,...,\pa_n\vf\Bigr\}^\top$, written in the full system of vectors  $\big\{\d\,^j\big\}_{j=1}^{n+1}$ in \eqref{e5.15} coincides with the
extended gradient $\cD_{\Omega^\ve}\,\vf$  in \eqref{e5.23}.

Similarly: the classical divergence $\Div\,\U:=\dst\sum\limits_{j=1}^n \pa_jU_j$ of a vector field $\U:=\dst\sum\limits_{j=1}^nU_j\e^j$, written in the full system \eqref{e5.15}, coincides with the extended divergence $\Div\,\U=\Div_{\Omega^\ve}\,\U$ in \eqref{e5.24}.

The extended gradient and the negative extended divergence are dual  $\nabla^*_{\Omega^\ve}=-\Div_{\Omega^\ve}$ and $\Div^*_{\Omega^\ve}=-\cD_{\Omega^\ve}$.

The Laplace-Beltrami operator $\Delta_{\Omega^\ve}:=\Div_{\Omega^\ve} \cD_{\Omega^\ve}\,\vf=-\nabla^*_{\Omega^\ve}\,\Big(\cD_{\Omega^\ve}\vf\Big)$ on $\Omega^\ve$, written in the full system \eqref{e5.15},  acquires the following form
 \begin{eqnarray}\label{e5.26}
\Delta_{\Omega^\ve}\vf=\sum\limits_{j=1}^{n+1}\cD_j^2\vf\, ,\quad
    \vf\in\bW^2(\Omega^{\ve})\, .
 \end{eqnarray}
 \end{lemma}
\noindent {\em Proof:} A similar lemma is proved in \cite[Lemma 4.3]{Du5}, but definition of the divergence $\Div_{\Omega^\ve}$ is different there. Therefore we expose the full proof below.

That the gradients coincide follows from the choice of the full system \eqref{e5.15}:
 \begin{eqnarray}\label{e5.27}
\nabla\vf&\hskip-3mm:=&\hskip-3mm\Bigl\{\pa_1\vf,...,\pa_n\vf
     \Bigr\}^\top=\sum_{j=1}^n(\pa_j\vf)\e^j=\sum_{j=1}^n
     (\cD_j\vf+\cN_j\cD_{n+1}\vf)\e^j\nonumber\\
&\hskip-3mm=&\hskip-3mm\sum_{j=1}^n(\cD_j\vf)\d\,^j+(\cD_{n+1}\vf)\cN
     =\sum_{j=1}^{n+1}(\cD_j\vf)\d\,^j=\cD_{\Omega^\ve}\vf
 \end{eqnarray}
since
 \begin{eqnarray}\label{e5.28}
 \begin{array}{c}
\e^j=\d^j+\cN_j\cN,\quad \pa_j=\cD_j+\cN_j\cN,\\
\dst\sum_{j=1}^n\cN_j\cD_j=0,\quad
     \dst\sum_{j=1}^n(\cD_j\vf)\e^j=\dst\sum_{j=1}^n(\cD_j\vf)\d\,^j.
 \end{array}
 \end{eqnarray}
By applying \eqref{e5.19} and \eqref{e5.28} we proceed as follows:
 \begin{eqnarray}\label{e5.29}
\Div\,\U&\hskip-3mm=&\hskip-3mm\sum_{j=1}^n\pa_jU_j=\sum_{j=1}^n\cD_jU_j
    +\sum_{j=1}^n\cN_j\pa_\cN U_j=\sum_{j=1}^n\cD_j\left[U^0_j+
    \cN_j\langle\cN,\U\rangle\right]\nonumber\\
&&+\sum_{j=1}^n\pa_\cN\big(\cN_jU_j\big)=\sum_{j=1}^n\cD_jU^0_j
    +\sum_{j=1}^n(\cD_j\cN_j)\langle\cN,\U\rangle
    +\cD_{n+1} U^0_{n+1}\nonumber\\
&\hskip-3mm=&\hskip-3mm\sum_{j=1}^{n+1}\cD_jU^0_j+\cH^0_\cC\langle\cN,\U\rangle
    =\Div_{\Omega^\ve}\U.
 \end{eqnarray}

The proved equality and the classical equality  $\nabla^*=-\Div$, ensure the both claimed equalities $\nabla^*_{\Omega^\ve}=-\Div_{\Omega^\ve}$ and $\Div^*_{\Omega^\ve}=-\cD_{\Omega^\ve}$:
 \[
\scal\cD_{\Omega^\ve}\vf,\U\scar=\scal\nabla\,\vf,\U\scar
     =-\scal\vf,\Div\,\U\scar=-\scal\vf,\Div_{\Omega^\ve}\U\scar.
 \]

Formula \eqref{e5.26} for the Laplace-Beltrami operator is a direct consequence of equalities \eqref{e5.27}, \eqref{e5.29} and definitions
 \[
\Dlb\vf=\Div\,\nabla\vf=\Div_{\Omega^\ve}\cD_{\Omega^\ve}\vf=\sum_{j=1}^{n+1}
     \cD^2_j\vf+\langle\cN,\cD_{\Omega^\ve}\vf\rangle
     =\sum_{j=1}^{n+1}\cD^2_j\vf,
 \]
because (see the third formula in \eqref{e5.28}) $\langle\cN,\cD_{\Omega^\ve}\vf\rangle=\dst\sum_{j=1}^n\cN_j\cD_j\vf=0$.     \QED

Let us check the following equalities for a later use:
 \begin{eqnarray}\label{e5.30}
\cD_{\Omega^\ve}\U=\big[\cD_jU^0_k\big]_{n+1\times n+1}+\langle\cN,\U\rangle\cW_{\Omega^\ve},
 \end{eqnarray}
where
 \begin{eqnarray*}
\U:=\dst\sum_{m=1}^{n+1}U^0_m\d\,^m=\dst\sum_{m=1}^nU_m\e^m,\quad
     U^0_{n+1}=\dst\sum_{m=1}^n\cN_mU_m,\quad \cD_{n+1}:=\pa_\cN,\quad \d^{n+1}:=\cN.
 \end{eqnarray*}
$\cW_{\Omega^\ve}$ is the extended Weingarten matrix (cf. \eqref{e2.12})
 \begin{eqnarray}\label{e5.31}
\cW_{\Omega^\ve}:=\big[\cD_j\cN_k\big]_{n+1\times n+1}
 \end{eqnarray}
and its last column and last row are $0$, because $\cD_j\cN_{n+1}=\cD_{n+1}\cN_j=\cD_{n+1}\cN_{n+1}=0$ for $j=1,\ldots,n$.

In fact (see \eqref{e5.27} fore some further details of calculation):
 \begin{eqnarray*}
\cD_{\Omega^\ve}\U&\hskip-3mm:=&\hskip-3mm
     \big[\pa_jU_k\big]_{n\times n}=\sum_{j,ky=1}^n\pa_jU_k\e^j\otimes\e^k\\
&\hskip-3mm:=&\hskip-3mm\sum_{j,k=1}^n[\cD_j+\cN_j\pa_\cN][U^0_k+\cN_k\langle
     \cN,\U\rangle][\d^j+\cN_j\cN]\otimes[\d^k+\cN_k\cN]\\
&\hskip-3mm=&\hskip-3mm\sum_{j,k=1}^n(\cD_jU^0_k)\d^j\otimes[\d^k+\cN_k\cN]
     +\sum_{j,k=1}^n\cD_j[\cN_k\langle\cN,\U\rangle]\d^j\otimes[\d^k+\cN_k\cN]\\
&&+\sum_{j,k=1}^n\cN^2_j(\pa_\cN U^0_k)\cN\otimes[\d^k+\cN_k\cN]
     +\sum_{j,k=1}^n\cN^2_j\cN^2_k\pa_\cN\langle\cN,\U\rangle\cN\otimes\cN
 \end{eqnarray*}
 \vskip-10mm
 \begin{eqnarray*}
&\hskip-3mm=&\hskip-3mm\sum_{j,k=1}^n(\cD_jU^0_k)\d^j\otimes\d^k
     +\sum_{j,k=1}^n\cN_k(\cD_jU^0_k)\d^j\otimes\d^{n+1}\\
&&+\sum_{j,k=1}^n\langle\cN,\U\rangle(\cD_j\cN_k)\d^j\otimes[\d^k+\cN_k\cN]
     +\sum_{j,k=1}^n\cN_k^2\cD_j\langle\cN,\U\rangle\d^j\otimes\d^{n+1}\\
&&+\sum_{k=1}^n(\cD_{n+1}U^0_k)\d^{n+1}\otimes\d^k
     +\sum_{k=1}^n\left[\cN_k\cD_{n+1}U^0_k
     +\cD_{n+1}U^0_{n+1}\right]\d^{n+1}\otimes\d^{n+1}\\
&\hskip-3mm=&\hskip-3mm\sum_{j,k=1}^n(\cD_jU^0_k)\d^j\otimes\d^k
     +\sum_{j,k=1}^n\left[\cD_j(\cN_kU^0_k)-U^0_k\cD_j\cN_k
     \right]\d^j\otimes\d^{n+1}\\
&&+\langle\cN,\U\rangle\sum_{j,k=1}^n(\cD_j\cN_k)\d^j\otimes\d^k
     +\sum_{j=1}^n\cD_j\langle\cN,\U\rangle\d^j\otimes\d^{n+1}\\
&&+\sum_{k=1}^n(\cD_{n+1}U^0_k)\d^{n+1}\otimes\d^k
     +(\cD_{n+1}U^0_{n+1})\d^{n+1}\otimes\d^{n+1}\\
&\hskip-3mm=&\hskip-3mm\sum_{j,k=1}^{n+1}(\cD_jU^0_k)\d^j\otimes\d^k
     -\sum_{j,k=1}^nU^0_k(\cD_j\cN_k)\d^j\otimes\d^{n+1}
     +\langle\cN,\U\rangle\sum_{j,k=1}^n(\cD_j\cN_k)\d^j\otimes\d^k\\
&\hskip-3mm=&\hskip-3mm\big[\cD_jU_k\big]_{(n+1)\times(n+1)}
     +\langle\cN,\U\rangle\cW_{\Omega^\varepsilon}
     -\sum_{j,k=1}^nU^0_k(\cD_j\cN_k)\d^j\otimes\d^{n+1}
 \end{eqnarray*}
 \begin{eqnarray*}
&\hskip-3mm=&\hskip-3mm\big[\cD_jU_k\big]_{(n+1)\times(n+1)}
     +\langle\cN,\U\rangle\cW_{\Omega^\varepsilon}
     -\big[(\cW_{\Omega^\varepsilon}\U^0)_j\delta_{j,n+1}\big]_{(n+1)\times(n+1)},
 \end{eqnarray*}
since
 \begin{eqnarray*}
\pa_\cN\cN_j=0,\quad\sum_{j,k=1}^n\cN^2_j=1,\quad \sum_{j=1}^n\cN_j\cD_j=0,
     \quad \sum_{j=1}^n\cN_j\d^j=0,\\
\sum_{k=1}^n\cN_kU^0_k=0,\quad\sum_{k=1}^n\cN_k\cD_j\cN_k=\frac12\cD_j\sum_{k=1}^n\cN^2_k=\frac12\cD_j1=0,
     \quad j=1,2\ldots,n+1.
 \end{eqnarray*}

 %
 %
\section{Convex energies}
 \label{sec5}
\setcounter{equation}{0}

Let again $\Omega^\varepsilon$ be a layer domain of width $2\varepsilon$ in the direction transversal to the mid-surface $\cC$ (see \S\, \ref{sec4}).

Any minimizer $u$ of the energy functional
 \begin{equation}\label{e6.1}
\cE^\varepsilon(u):=\int_{\Omega^\varepsilon}\langle\nabla\,u,\nabla\,u
     \rangle\,dy, \qquad u\in C^\infty(\Omega^\varepsilon)
 \end{equation}
should satisfy
 \begin{eqnarray}\label{e6.2}
0&\hskip-3mm=&\hskip-3mm\frac{d}{dt}\cE^\varepsilon(u+tv)\Bigl|_{t=0}
     =\int_{\Omega^\varepsilon}\left[\langle\nabla\,u,\nabla\,v\rangle
     +\langle\nabla\,v,\nabla\,u\rangle\right]\,dy \nonumber\\
&\hskip-3mm=&\hskip-3mm2{\rm Re}\int_{\Omega^\varepsilon}\langle\nabla\,u,
     \nabla\,v\rangle\,dy=-2{\rm Re}\int_{\Omega^\varepsilon}\langle\Div\nabla\,
     u,v\rangle\,dy=-2{\rm Re}\int_{\Omega^\varepsilon}\langle\Delta\,
     u,v\rangle\,dy
 \end{eqnarray}
for arbitrary $u\in C^\infty(\Omega^\varepsilon)$ and $v\in C^\infty_0(\Omega^\varepsilon)$, which implies
 \begin{equation}\label{e6.3}
\Delta\,u=0\qquad\mbox{ on }\quad \Omega^\varepsilon.
 \end{equation}

In other words, \eqref{e6.3} is the Euler-Lagrange equation associated with the energy functional \eqref{e6.1}.

Similarly, minimizers of the energy functional
 \begin{equation}\label{e6.4}
\cE_0(u):=\int_{\cC}\langle\nabla_\cC u,\nabla_\cC u\rangle\,d\sigma, \qquad
     u\in C^\infty(\cC)
 \end{equation}
on the hypersurface $\cC$ should satisfy the following Laplace-Beltrami equation
 \begin{equation}\label{e6.5}
\Delta_\cC u:=\Div_\cC\nabla_\cC u=0\qquad\text{on}\quad \cC.
 \end{equation}
To treat the dimension reduction problem for the Laplace equation (see \cite{Br1} for a similar consideration in case of a flat 3D body), we
assume, without restricting generality, that $\Omega^1$ (i.e., for $\varepsilon=1$) is still a layer domain. Otherwise we can first change the variable $\cx_n=\ve_0\bar\cx_n$, $0<\bar\cx_n<1$, where $0<\ve_0<1$ is such that $\Omega^{\varepsilon_0}$ is still a layer domain.

Next we introduce a new coordinate system (cf. \eqref{e5.19})
 \begin{eqnarray}\label{e6.6}
 \begin{array}{c}
x:=\dst\sum_{m=1}^nx_m\e^m=\dst\sum_{m=1}^n\cx_m\d\,^m+t\d\,^{n+1},\\
\cx_k:=x_k-\cN_k\langle\cN,x\rangle,\quad k=1,\ldots,n,\quad t=\cx_{n+1}
     :=\langle x,\cN\rangle=\dst\sum_{m=1}^nx_m\cN_m
 \end{array}
 \end{eqnarray}
and the scalar product of elements
 \[
\cx:=\dst\sum_{m=1}^{n+1}\cx_m\d\,^m,\quad \cy:=\dst\sum_{m=1}^{n+1}\cy_m\d\,^m
 \]
define by the equality (cf. similar in \eqref{e5.32})
 \[
\langle\cx,\cy\rangle:=\sum_{j=1}^{n+1}\cx_j\overline{\cy_j}.
 \]
Then (cf. \eqref{e5.33}-\eqref{e5.34})
 \begin{eqnarray}\label{e6.7}
\langle\cx,\cy\rangle=\sum_{j=1}^{n+1}\cx_j\overline{\cy_j}
     =\sum_{j=1}^n(x_j-\cN_j\langle\cN,x\rangle)(\overline{(y_j-\cN_j\langle
     \cN,y\rangle)})+\langle\cN,x\rangle\langle\cN,y\rangle\nonumber\\
=\sum_{j=1}^nx_j\overline{y_j}=\langle x,y\rangle.
 \end{eqnarray}
In particular,
 \[
\|\cx\|:=\sum_{j=1}^{n+1}|\cx_j|^2=\sum_{j=1}^n|x_j|^2=\|x\|.
 \]

Due to Lemma \ref{l5.7} the classical gradient in the energy functional \eqref{e6.1} can be replaced by the extended gradient
 \begin{eqnarray}\label{e6.9}
\cE^\varepsilon(u)&\hskip-3mm:=&\hskip-3mm\int_{\Omega^\varepsilon}\langle
     \cD_{\Omega^\varepsilon}u(y),\cD_{\Omega^\varepsilon} u(y)\rangle\,dy\nonumber\\
&\hskip-3mm=&\hskip-3mm\int_{-\varepsilon}^\varepsilon
     \int_\cC\left[\langle\cD_\cC u(\cx,t),\cD_\cC u(\cx,t)\rangle
     +|\pa_tu(\cx,t)|^2\right]\,d\sigma\,dt,\\
&&\hskip50mm\cD_\cC:=\left(\cD_1,\ldots,\cD_n\right)^\top\nonumber
 \end{eqnarray}
for arbitrary $u\in\bW^1(\Omega^\varepsilon)$, because $\cD_{n+1}=\pa_\cN =\pa_t$. Here $\cC$ is the mid surface of the layer domain $\Omega^\varepsilon=\cC\times(-\varepsilon,\varepsilon)$ and $d\sigma$ is the surface measure on $\cC$.

Due to the representation \eqref{e6.9} and the new coordinate system \eqref{e6.6} we can apply the scaling with respect to the variable $t$ and study the scaled energy. The approach is based on $\Gamma$-convergence (see \cite{Br1,FJM1}) and can be applied to a general energy functional which is convex and has square growth. The problem we have in mind is the following: \emph{Do these energies defined on thin n-dimensional domains $\Omega^\varepsilon$ converge (and in which sense) to an energy defined on the $n-1$ dimensional Hypersurface $\cC$ (the mid-surface of $\Omega^\varepsilon$) when the domain $\Omega^\varepsilon$ is "squeezed" infinitely in the transversal direction to $\cC$?}

In the next two sections we apply the results developed in the present paper to boundary value problems for the heat conduction by a hypersurface. In particular we shall show, that when the thickness of the layer domain $\Omega^\varepsilon$, with the mid-surface $\cC$, tends to zero, a solution to the linear heat conduction equation Gamma-converges to a solution to the certain boundary value problem Laplace-Beltrami equation on the mid-surface $\cC$ written explicitly (see \ref{sec4}).

 %
 %
\section{Variational reformulation of a heat transfer problems}
\label{sec6}
\setcounter{equation}{0}

Let $\Omega$ be a bounded Lipschitz domain in $\mathbb{R}^3$ with piecewise smooth boundary $\partial \Omega = \overline{\cC}_D \cup \overline{\cC}_N$, where $\cC_D$ and $\cC_N$ are open non-intersecting surfaces $\cC_D\cap\cC_N= \varnothing$ and their common boundary is a smooth arc. Denote by $\nub=(\nu_1,\nu_2,\nu_3)^\top$ the unit normal on $\cC$, external with respect to $\Omega$.

We consider the general steady-state, linear heat transfer problem for a medium occupying  domain $\Omega$. We assume that on the $\cC_D$ part of the boundary $\partial \Omega$ temperature $g$ is prescribed, while on the $\cC_N$ part of $\partial \Omega$ is prescribed heat flux $q$.

We look for a temperature distribution $T(x)$ in $\Omega$, which satisfies the linear heat conduction equation
\begin{equation}\label{ht_1}
\Div(\cA(x)\nabla T)(x)=f(x), \qquad x\in \Omega
\end{equation}
and boundary conditions
\begin{eqnarray}\label{ht_2}
& T^+(y)=g(y)\qquad {\rm on}  \;\; \cC_D,\\
\label{ht_3}
&-\langle\nub(y),\cA^+(y)(\nabla T)^+(y)\rangle=q(y)\qquad{\rm on}\;\; \cC_N,
\end{eqnarray}
where $\cA$ is the thermal conductivity, $f$ is the heat source, $g$ is the distribution of temperature and $q$ is the heat flux, which are supposed known.

We will suppose, that $\cA(x)$ is a continuous $3\times3$ matrix-function, positive definite in the following sense (see \eqref{e3.26}), which implies the inequality
 \begin{eqnarray}\label{ht_02}
\scal\cA\U,\U\scar\geqslant C\|\U|\bL_2(\Omega)\|^2
 \end{eqnarray}
valid for all 3-vectors $\U=(U_1,U_2,U_3)^\top\in\bL_2(\Omega)$. The conditions on $\cA$ imply that the traces $\cA^+(y)$ at the boundary $\cC$ exist and $\cA^+$ has the same properties, namely, is a continuous positive definite matrix function.

We impose the following natural constraints on the solution $T$ and functions $f$, $g$ and $q$, which are prescribed:
 \begin{eqnarray}\label{ht_0}
T\in\bH^1(\Omega),\quad f\in\wt\bH^{-1}(\Omega), \quad
     g\in\bH^{1/2}(\cC_D),\quad  q\in\bH^{-1/2}(\cC_N).
 \end{eqnarray}
The existence of the traces $\langle\nub(y),\cA^+(y)(\nabla T)^+\rangle \in\bH^{-1/2} (\cC_3)$, which is not ensured by the trace theorem, follows from the Green formula
 \begin{eqnarray}\label{ht_03}
\int_\Omega(\Div\,\cA(x)\nabla T)(x)\psi(x)dx=\int_\cC\langle\nub(y),\cA^+(y)
     (\nabla T)^+(y)\rangle\psi^+(y)\,d\sigma\nonumber\\
-\int_\Omega\langle \cA(x)\nabla T(x),\nabla\psi(x)\rangle\,dx
 \end{eqnarray}
by the duality between the spaces $\bH^{1/2}(\cC)$ and $\bH^{-1/2}(\cC)$ due to the fact that $T$ is a solution to the equation \eqref{ht_1}. For this  we rewrite \eqref{ht_03} in the form
 \[
\int_\cC\langle\nub(y),\cA^+(y)(\nabla T)^+(y)\rangle\psi^+(y)\,d\sigma
     =\int_\Omega f(x)\psi(x)dx+
     \int_\Omega\langle \cA(x)\nabla T(x),\nabla\psi(x)\rangle\,dx,
 \]
and note that $\psi\in\bH^1(\Omega)$ is arbitrary and, therefore,
$\psi^+\in\bH^{1/2}(\cC)$ is arbitrary.

Let $\Omega\subset\bR^n$ be a domain with a Lipshitz boundary $\cM:=\partial\Omega$ and $\cM_0\subset\partial\Omega$-be a subsurface of the boundary surface which has the non-zero measure. By $\wt\bH^1(\Omega,\cM_0)$ we denote a subspace of $\wt\bH^1(\Omega)$ of those functions which have vanishing traces on the part of the boundary
\begin{eqnarray}\label{ht_06}
\wt\bH^1(\Omega,\cM_0):=\left\{\vf\in\bH^1(\Omega)\;:\;
     \vf^+(y)=0\quad\forall\, y\in\cM_0\right\}.
 \end{eqnarray}
This space inherits the standard norm from $\bH^1(\Omega)$:
\begin{eqnarray*}
\|\vf\,\big|\,\bH^1(\Omega)\,\|:&=\left[\|\varphi\,\big|\,\bL(\Omega)\,\|^2
     +\sum\limits_{j=1}^n\|\pa_j\vf\,\big|\,\bL_2(\Omega)\|^2\right]^{1/2}.
\end{eqnarray*}

Consider the functional
\begin{eqnarray}\label{ht_5}
\Phi (T)=\int\limits_{\Omega}\left[\frac12\langle\cA(x)\nabla T(x),\nabla T(x)\rangle
     +f(x)T(x)\right]dx+\int\limits_{\cC_N}q(y) T^+(y)d\sigma
 \end{eqnarray}
where $f$ and $q$ satisfy conditions \eqref{ht_0} and $T\in\bH^1(\Omega)$ has vanishing traces on $\cC_D$, i.e., $T\in\wt\bH^1(\Omega,\cC_D)$ (see \eqref{ht_06}).

The second summand in the in integral on $\Omega$ is understood in the sense of duality between the spaces $\wt\bH^{-1}(\Omega)$ and $\bH^1(\Omega)$. Concerning the integral on $\cC_N$: it is understood in the sense of duality between the spaces  $\wt\bH^{1/2}(\cC_N)$ and $\bH^{-1/2}(\cC_N)$ because $q\in\bH^{-1/2}(\cC_N)$ and, due to the condition inclusion $T\in\wt\bH^1(\Omega,\cC_D)$, $\supp\,T^+\subset\cC_N$ which implies $T^+\in\wt\bH^{1/2}(\cC_N)$.
 %
 \begin{theorem}\label{t6.1}
The problem \eqref{ht_1}-\eqref{ht_3} with vanishing Dirichlet condition $T^+(y)=g(y)=0$ for all $y\in\cC_D$ is reformulated into the following equivalent variational problem: Let $f$ and $q$ satisfy conditions \eqref{ht_0} and look for a temperature distribution $T\in\wt\bH^1(\Omega,\cC_D)$ (see \eqref{ht_06}) which is a stationary point of the functional \eqref{ht_5}.
 \end{theorem}
{\bf Proof:} Let $T(x)$ be a stationary point of the functional \eqref{ht_5}, where $\Phi(T)$ attains a local infimum. Consider  the variation
\begin{eqnarray}\label{e6.8}
\delta\Phi=\frac{d}{d\varepsilon}\Phi(T+\varepsilon\V)|_{\varepsilon=0}
     =\int\limits_{\Omega}\big[\langle \cA(x)\nabla T(x),\nabla\V(x)\rangle+f(x)\V(x)\big] dx\nonumber\\
+\int\limits_{\cC_N}q(y)\V^+(y).
 \end{eqnarray}
The trial function $\V\in\bH^1(\Omega)$ is such that $T+\varepsilon\V$ satisfies the boundary conditions. Then from the equalities $T^+(y)+\V^+(y)=0=T^+(y)$ on $\cC_D$ follows that $T^+(y)=\V^+(y)=0$ on $\cC_D$, i.e., $T$ and $\V$ have the traces zero on the part $\cC_D$ of the boundary.

It is clear, that for those $\V$ for which the functional $\Phi(T+\ve\V)$ attains infimum, we have $\delta\Phi=0$. By applying the Gau\ss \ theorem to the first summand under the integral on $\Omega$ in \eqref{e6.8}, we obtain the associated Euler-Lagrange equation
\begin{eqnarray}\label{ht_7}
\int\limits_{\Omega}\big[-\Div \cA(x)\nabla T(x) + f(x)\big]
     \V(x)\,dx + \int\limits_{\cC_D}\langle \nub(y),\cA^+(y)(\nabla T)^+(y)\rangle\V^+(y) d\sigma\nonumber\\
+\int\limits_{\cC_N}\Big[q(y)+\langle\nub(y),\cA^+(y)(\nabla T)^+(y)
     \rangle\Big]\V^+(y)d\sigma=0.
 \end{eqnarray}

Since the trial function $\V$ vanishes on $\cC_D$ (see \eqref{ht_06}), the integral on $\cC_D$ in \eqref{ht_7} vanishes. Now taking arbitrary function $\V\in C_0^{\infty}(\Omega)$ (vanishing in the vicinity of the boundary $\cC$), all summands in \eqref{ht_7} except the first one vanish and we obtain
\begin{equation}\label{ht_8}
\int\limits_\Omega \big[-\Div \cA(x)\nabla T(x) + f(x)\big]\V(x)\,dx=0,
 \end{equation}
which is equivalent to the basic differential equation in \eqref{ht_1}.

Therefore from \eqref{ht_7} follows that
\begin{eqnarray}\label{ht_9}
&&\int\limits_{\cC_N}\Big[q(y)+\langle\nub(y),\cA^+(y)(\nabla T)^+(y)\rangle
     \Big]\V^+(y)\,d\sigma=0.
\end{eqnarray}
The trace  $\V^+$ of a trial function in \eqref{ht_9} is arbitrary, we derive, the boundary condition \eqref{ht_3}.

Vice versa: Let $T$ be a solution to the mixed problem \eqref{ht_1}-\eqref{ht_3} with vanishing Dirichlet traces $T^+(y)=g(y)=0$ on $\cC$, by taking the scalar product of the basic equation in \eqref{ht_1} with the solution $T$, by applying the Green formulae and the boundary conditions \eqref{ht_2} with $g=0$, we get the following equality:
\begin{eqnarray*}
0&\hskip-3mm=&\hskip-3mm\int\limits_{\Omega}\big[-\Div \cA(x)\nabla T(x)
     + f(x)\big] T(x)\,dx=\int\limits_{\Omega}\big[\cA(x)\nabla T(x) + f(x)\big]\nabla T(x)\,dx\nonumber\\
&&+\int\limits_{\cC_D\cup\cC_N}\langle \nub(y),\cA^+(y)(\nabla T)^+(y)\rangle T^+(y)
     d\sigma\nonumber\\
&\hskip-3mm=&\hskip-3mm\int\limits_{\Omega}\big[\cA(x)\nabla T(x) + f(x)\big]\nabla
     T(x)\,dx\int\limits_{\cC_N}q(y)T^+(y)d\sigma.
 \end{eqnarray*}
Therefore, $T$ is a stationary point of the functional $\Phi$ in \eqref{ht_5}.    \QED
 %
 \begin{corollary}\label{c.cor}
The minimization problem for the functional \eqref{e7.38} is an equivalent reformulation of the BVP \eqref{e7.37}.
 \end{corollary}

If $\cC_D=\cC, \; \cC_N=\emptyset$,  the problem \eqref{ht_1}-\eqref{ht_3} reduces to the problem with a Dirichlet boundary condition
\begin{equation}
\label{ht_10}
 T^+(y)=0 \qquad {\rm on}\quad \cC
\end{equation}
and the corresponding functional $\Phi$ in variational formulation (see \eqref{ht_5}) takes the form
\begin{equation}
\label{ht_11}
\Phi_D(T)=\frac12\int\limits_{\Omega}\Big[\langle \cA(x)\nabla T(x),\nabla T(x)\rangle+ f(x)T(x)\Big]\,dx.
 \end{equation}
If $\cC_D=\emptyset, \; \cC_N=\cC$, from \eqref{ht_1}-\eqref{ht_3} we get the problem with Neumann boundary condition
\begin{equation}
\label{ht_12}
-\langle \cA^+(y)\nub(y),(\nabla T)^+(y)=q(y) \qquad {\rm on}\quad \cC
\end{equation}
and the corresponding functional in variational formulation (see \eqref{ht_5}) takes the form
\begin{equation}
\label{ht_13}
\Phi_N(T)=\frac12\int\limits_{\Omega}\Big[\langle \cA(x)\nabla T(x),\nabla T(x)
     \rangle + f(x)T(x)\Big]\,dx + \int\limits_\cC q(y)T^+(y)d\sigma.
 \end{equation}

 %
 %
\section{Heat transfer in thin Layers}
 \label{sec7}
\setcounter{equation}{0}

Let $\cC$ be a $C^2$ smooth orientable  surface in $\mathbb{R}^3$ given by a single chart (immersion)
$$
\theta\;:\;\omega \rightarrow \cC, \qquad \omega \subset \mathbb{R}^2
$$
and let $\nu(\cx),\; \cx \in \cC$ be the unit normal on $\cC$ with the chosen orientation. Chart is supposed to be single just for convenience and multi-chart case can be considered similarly. Denote by $\Omega^\varepsilon $ the layer domain i.e. the set of all points in $\mathbb{R}^3$ in the distance less then $\varepsilon$ from $\cC$. Then for sufficiently small $\varepsilon$ the map $\Theta\;:\; \cC \times (-\varepsilon, \varepsilon)\rightarrow \Omega^\varepsilon $
\begin{equation}\label{ht_16}
\Theta(\cx,t)=\cx+t\nu(\cx)=\theta(x)+t\nu(\theta(x)), \qquad x\in \omega
\end{equation}
is $C^1$ homeomorphism and $\Theta(\cC \times \{0\})=\cC$.

As noted above we can properly extend normal field on the entire $\Omega^\varepsilon $ assuming
\begin{equation}\label{ht_16_1}
\nub(\cx+t\nub(\cx))=\nub(\cx),\qquad \cx\in\cC, \qquad -\varepsilon<t<\varepsilon.
\end{equation}
If $\varepsilon$ is sufficiently small, the boundary $\cM^\varepsilon:=\partial \Omega^\varepsilon $ is represented as the union of three $C^1$-smooth surfaces $\cM^\ve=\cM_{\ve,D}\cup\cM_{\ve,N}^{-}\cup\cM_{\ve,N}^{+}$, where $\cM_{\ve,D}=\partial \cC\times[-\varepsilon,\varepsilon]$ is the lateral surface, $\cM^+_{\ve,N}=\cC\times \{+\varepsilon\}$ is the upper surface and $\cM^-_{\ve,N}=\cC\times\{-\varepsilon\}$ is the lower surface of the of the boundary $\cM^\ve$ of layer domain $\Omega^\ve$.

In the present section we will consider heat conduction by an "isotropic" media, governed by the Laplace equations (the case $\cA(x)\equiv1$ in \eqref{ht_1}-\eqref{ht_3}). The case of an "anisotropic" media will be treated in a forthcoming publication in a thin layer domain $\Omega^\ve:=\cC\times(-\ve,\ve)=\Theta^{-1}(\Omega^\ve)$:
\begin{eqnarray}\label{ht_17}
\begin{array}{ll}
\Delta_{\Omega^\ve} T(\cx,t)=f(\cx,t), \qquad &(\cx,t)\in
     \cC\times(-\ve,\ve),\\[3mm]
T^+(\cx,t)= 0,  & (\cx,t)\in \pa\cC\times(-\ve,\ve),\\[3mm]
\pm(\pa_t T)^+(\cx,\pm\ve)=q^\pm(\cx), & \cx\in\cC,
\end{array}
\end{eqnarray}
where (see \eqref{e5.26}, \eqref{e6.5})
$$
\Delta_{\Omega^\ve}T=\sum\limits_{j=1}^{4}\cD_j^2 T=\Delta_{\cC}T+\pa_t^2 T.
$$
The different signs $\pm(\pa_t T)^+(\cx,\pm\ve)$ in the third equality in \eqref{ht_17} is due to the different orientation of the outer unit normal vector $\nub(\cx)$ at the upper and lower surfaces $\cC\times\{\pm\ve\}$.

We impose the following constraints
 \begin{eqnarray}\label{ht_171}
 \begin{array}{c}
 T\in\bH^{1}(\Omega^{\varepsilon}),\qquad f\in\bL_2(\Omega^1),\\
 0 \;\; \text{is the Lebesgue point for the function}\;\; F(t):=\dst\int_\cC |f(\cx,t)|^2d\sigma
 \end{array}
 \end{eqnarray}
(see \eqref{e3.32b} and note that $\|F\big|\bL_1(-1,1)\|\leqslant\|f\big| \bL_2(\Omega^1)\|^2$). The latter constraint implies that $F(0)$ exists and, due to Theorem \ref{t7.7},
 \begin{eqnarray}\label{e7.7}
 \frac1\ve \int_{-\ve}^\ve F(t)dt=\frac1\ve\int_{-\ve}^\ve\int_\cC|f(\cx,t)|^2d\sigma dt
 \leqslant 2F(0)<\infty
 \end{eqnarray}
for all $0<\ve<\ve_0$ and some small $\ve_0>0$ (cf. the definition of a Lebesgue point \eqref{e3.32b}).

Conditions \eqref{ht_171} are slightly more restrictive on $f$ than is necessary for the solvability (see \eqref{ht_0}) and is needed for the $\Gamma$-convergence.

The next example demonstrates that not all functions in $\bL_2(\Omega^1)$ have the property \eqref{ht_171}. Let
\begin{equation*}
f\left( \cx,t\right)=\begin{cases}
\sqrt{\left( -\dst\frac1{\ln t}\right)^\prime}=\dst\frac1{t^{1/2}\log t},\qquad
     &\text{for}\quad t\in\left(0,\dst\frac12\right),\\
0, &\text{for}\quad t\notin \left( 0,\dst\frac12\right).
\end{cases}
\end{equation*}

It is easy to show that%
\begin{eqnarray*}
\left\Vert f\left( \cx,t\right) \big|\,\bL_{2}(\Omega ^{1})\right\Vert
&=&\int_{-1}^{1}\int_{\cC}|f(\cx,t)|^{2}d\sigma dt=\int_{\cC}d\sigma
\int_{0}^{1/2}\frac{dt}{t\ln^2t} \\
&=&-C\int_{0}^{1/2}\left(\frac1{\ln t}\right)^\prime dt=-\left.\frac C{\ln t}\right\vert_0^{1/2}=\frac C{\ln 2}<\infty.
\end{eqnarray*}

On the other hand, if $F(t)$ is defined in \eqref{ht_171},
\begin{eqnarray*}
\frac{1}{\ve}\int_{-\ve}^{\ve}F(t)dt &=&\frac{1}{\ve}\int_{-\ve}^{\ve}\int_{\cC}
|f(\cx,t)|^{2}d\sigma dt=\int_{\cC}d\sigma \frac{1}{\ve}\int_{0}^{\ve}
\frac{dt}{t\ln^2t} \\
&=&-\frac{C}{\ve}\int_{0}^{\ve}\left( \frac1{\ln t}\right)^\prime dt=-\left.\frac C
     {\ve\ln t}\right\vert_0^\ve=-\frac C{\ve\ln\ve}
     \rightarrow\infty,\quad \text{as}\quad\ve\rightarrow 0
\end{eqnarray*}
and $0$ is not the Lebesgue point for the function $F(t)$.
 %
\begin{remark}\label{r7.5}
Note, taking the Dirichlet and the Neumann traces zero $T^+(\cx,t)=0$ on $\pa\cC\times(-\ve,\ve)$ and on $\cC\times\{\pm\ve\}$, (see \eqref{ht_17}) we need to prove the $\Gamma$-convergence (see the Remark \ref{r0.2} above;

On the other hand, a BVP
\begin{eqnarray}\label{ht_17_0}
\Delta_{\Omega^\ve}T_0(\cx,t)&\hskip-3mm=&\hskip-3mmf(\cx,t), \hskip18mm
     (\cx,t)\in\cC\times(-\ve,\ve),\\
\label{ht_18_0}
T_0^+(\cx,t)&\hskip-3mm=&\hskip-3mm g(\cx,t), \hskip18mm (\cx,t)\in \pa\cC\times(-\ve,\ve),\\
\label{ht_19_0}
\pm(\pa_t T_0)^+(\cx,\pm\ve)&\hskip-3mm=&\hskip-3mm q^\pm(\cx), \hskip25mm \cx \in \cC\times\{\pm\ve\}
\end{eqnarray}
with the non-zero Dirichlet and Neumann traces on the boundary and the standard constraints \eqref{ht_171} reduces to the equivalent BVP \eqref{ht_17}.

Indeed, let $G\in\bH^1(\Omega^\ve)$ be a solution to the Mixed boundary value problem
\begin{equation}\label{cc1}
\begin{array}{ll}
\Delta_{\Omega^\ve}G(x)=0, &x\in\Omega^\ve,\\[2mm]
G^+(\cx)=g(\cx), &(\cx,t)\in \pa\cC\times(-\ve,\ve),\\[2mm]
\pm(\pa_t G)^\pm(\cx,\ve)=q^\pm(\cx), \qquad &\cx \in \cC\times\{\pm\ve\}.
\end{array}
\end{equation}
The unique solvability of the problem \eqref{cc1} is a classical result and follows, for example, from the Lax-Milgram Lemma.

Then the difference $T:=T_0-G$ solves the BVP \eqref{ht_17}.
\end{remark}

The formulated BVP \eqref{ht_17} governs a heat transfer in the body $\Omega^\varepsilon$ when there are thermal sources or sinks in $\Omega^\varepsilon $. The temperature on the lateral surface $\pa\cC\times(-\ve,\ve)$ is zero and heat fluxes are equal and fixed on the upper and lover $\cC^\pm:=\cC\times\{\pm\ve)$ surfaces. It is well known, that the boundary value problem \eqref{ht_17} as well as it's equivalent problem \eqref{ht_17_0}-\eqref{ht_19_0} have the unique solution $T\in \mathbb{H}^1(\Omega^\varepsilon)$ (respectively, $T_0\in \mathbb{H}^1(\Omega^\varepsilon)$; see, e.g., \cite{DTT1}).

The energy functional associated with the problem \eqref{ht_17_0} - \eqref{ht_19_0} reads (cf. Theorem \ref{t6.1})
\begin{eqnarray}\label{ht_5x}
E(T)&\hskip-3mm=&\hskip-3mm\int\limits_{\Omega^\ve}\Big[\frac12\langle(\cD_{\Omega^\ve}T)(x),
     (\cD_{\Omega^\ve}T)(x)\rangle+f(x)T(x)\Big]dx.
 \end{eqnarray}
To justify the equality \eqref{ht_5x}, we remind that  expressing the Cartesian derivatives by means of  G\"unter's derivatives, according to \eqref{e5.26} we get
\begin{eqnarray*}
\langle\nabla T,\nabla T\rangle=\sum\limits_{j=1}^3|\partial_j T|^2
     =\sum\limits_{j=1}^4|\cD_j T|^2=\sum\limits_{j=1}^3 |\cD_j T|^2+|\partial_\nub T|^2=\langle\cD_{\Omega^\ve}T,\cD_{\Omega^\ve}T\rangle.
\end{eqnarray*}
More generally, we consider the non-linear functional
\begin{equation}\label{ht_21}
E(T)=\int\limits_{\Omega^\varepsilon}\cK((\cD_{\Omega^\ve}T)(x),\,T(x))\,dx,
 \end{equation}
 in the case of the functional \eqref{ht_5x} we have
\begin{eqnarray}\label{ht_22}
&&\hskip-22mm\cK(\cD_{\Omega^\ve}T,\,T)=
\frac12\langle\cD_{\Omega^\ve}T(\cx,t),\cD_{\Omega^\ve}T(\cx,t)\rangle+f(\cx,t)\,T(\cx,t).
\end{eqnarray}
 %
\begin{lemma}\label{l7.1}
Let $\Omega$ be a domain in $\mathbb{R}^n$ with the Lipshitz boundary $\cM:=\partial\Omega$ and $\cM_0\subset\cM$ be a subsurface of non-zero measure. Then the inequality
\begin{equation}\label{Poinc}
\|\varphi\,\big|\,\bL_2(\Omega)\|\leqslant C\|\nabla\varphi\,\big|\,\bL_2(\Omega)\|
     =C\left[\sum\limits_{j=1}^n\|\pa_j\varphi\,\big|\,\bL_2(\Omega)\|^2\right]^{1/2}
\end{equation}
holds for all functions $\varphi\in \wt{\bH}{}^1(\Omega,\cM_0)$ and the constant $C$ is independent of $\varphi$.

Moreover, Let $\cC\subset\mathbb{R}^n$ be a smooth hypersurface with the Lipschitz boundary $\Gamma=\partial\cC$, $\Omega=\cC\times[a,b]$ is a cylinder with the base $\cC$  and $\cM_0:=\Gamma_0\times[a,b]$, $\Gamma_0\subset\Gamma$. Then for all functions $\varphi\in \wt{\bH}{}^1(\Omega,\cM_0)$  the inequality
\begin{equation}\label{Poinc_2}
\|\varphi\,\big|\,\bL_2(\Omega)\|\leqslant C'\|\nabla_\cC\varphi\,\big|
     \,\bL_2(\Omega)\|=C'\left[\sum\limits_{j=1}^n\|\cD_j\varphi\,\big|\, \bL_2(\Omega)\|^2\right]^{1/2}
\end{equation}
holds with only surface gradient $\nabla_\cC:=(\cD_1,\ldots,\cD_n)^\top$ and the constant $C'$ is independent of $\varphi$.
 \end{lemma}
{\bf Proof:} The formula
\begin{eqnarray}\label{ht_7.124}
\|\varphi\,\big|\,\wt{\bH}^1(\Omega,\cM_0)\,\|
     :=\|\nabla\varphi\,\big|\,\bL_2(\Omega)\|
\end{eqnarray}
defines an equivalent norm in the space $\wt{\bH}{}^1(\Omega,\cM_0)$. Indeed, the inequality $\|\vf\,\big|\,\bH^1(\Omega,\cM_0)\,\|$ $\leqslant\|\vf\, \big|\,\bH^1(\Omega)\|$ with the standard norm $\|\vf\,\big|\,\bH^1(\Omega)\|$ on $\wt{\bH}{}^1(\Omega,\cM_0)$ is trivial. On the other hand, $\|\varphi\,\big|\,\wt{\bH}^1(\Omega,\cM_0)\,\|$ has all properties of a norm. Since other properties are trivial to check, we will only check that  $\|\varphi\,\big|\,\wt{\bH}^1(\Omega,\cM_0)\,\|=\|\nabla\varphi\,\big|\,\bL_2(\Omega)\,\|=0$ implies $\varphi=0$. Indeed, the trivial norm implies that the gradient vanishes $\nabla\varphi=0$, which means that the corresponding function is constant $\varphi={\rm const}$; since $\varphi=0$ on $\cM_0$, it follows $\varphi\equiv0$.

If we apply the open mapping theorem of Banach (see \cite[Theorem 2.11, Corollary 2.12.b]{Ru73}, we conclude that the inverse inequality
 \[
\|\vf\,\big|\,\bH^1(\Omega)\|\leqslant C_1\|\vf\,\big|\,\bH^1(\Omega,\cM_0)\,\|
     =C_1\|\nabla\varphi\,\big|\,\bL_2(\Omega)\,\|
 \]
holds with some constant $C_1<\infty$. Since
 \[
\|\vf\,\big|\,\bH^1(\Omega)\|^2=\|\varphi\,\big|\,\bL_2(\Omega)\,\|^2
     +\|\nabla\varphi\,\big|\,\bL_2(\Omega)\,\|^2\leqslant C^2_1\|\nabla\varphi\,\big|\,\bL_2(\Omega)\,\|^2,
 \]
The claimed inequality \eqref{Poinc} follows with the constant $C:=\sqrt{C_1^2-1}$.

Now if $\cC$ is a hypersurface, $\Omega=\cC\times[a,b]$ is a cylinder with the base $\cC$ and $\cM_0:=\Gamma_0\times[a,b]$, $\Gamma_0\subset\Gamma :=\partial\cC$, in the space $\wt{\bH}{}^1(\Omega,\cM_0)$ we consider the semi-norm
\begin{equation}\label{Poinc_3}
\|\varphi\,\big|\,\bH^1(\Omega,\cM_0)\|:=\|\nabla_\cC\varphi\,\big|\,\bL_2(\Omega)\|
     =\left[\sum\limits_{j=1}^n\|\cD_j\varphi\,\big|\,\bL_2(\Omega)\|^2
     \right]^{1/2},
\end{equation}
which turns out to be a norm. Indeed, from $\|\varphi\,\big|\,\bH^1(\Omega,\cM_0)\|=\|\nabla_\cC\varphi\,\big|\,\bL_2
(\Omega)\|=0$ follows that $\varphi(\cx,t)=\varphi(t)$ is independent of the variable $\cx\in\cC$. Since $\varphi(t)=\varphi(\cx,t)=0$ for all $\cx\in\Gamma_0$ and all $t\in(a,b)$, it follows $\varphi\equiv0$. The proof is accomplished as in the foregoing case.    \QED
 %
\begin{lemma}\label{l7.2}
If $f\in\mathbb{L}_2(\Omega^1)$, $q^\pm_0\in\bH^{-1/2}(\cC)$, then the energy functional $E(T)$ in \eqref{ht_5x}-\eqref{ht_22} is correctly defined on the space $\wt{\bH}^1(\Omega^{\ve},\pa\cC\times(-\ve,\ve))$, is strictly convex and has the following quadratic estimate
\begin{eqnarray}\label{ht_121}
\begin{array}{l}
E(t T_1+(1-t)T_2)\leqslant t E(T_1)+(1-t)E(T_2),\\[3mm]
C_1\dst\int\limits_{\Omega^\ve}|(\cD_{\Omega^\ve}T)(x)|^2dx-C_2\leqslant E(T)\leqslant C_3\left[1 + \dst\int\limits_{\Omega^\ve}|(\cD_{\Omega^\ve}T)(x)|^2dx\right],\\
\hskip20mm\forall\,T_1,\ T_2\in\mathbb{H}^1(\Omega^\ve),\qquad \forall\,T \in\wt\bH^1(\Omega^{\ve},\pa\cC\times(-\ve,\ve))
\end{array}
 \end{eqnarray}
for some positive constants $C_1,C_2$ and $C_3$.
 \end{lemma}
{\em Proof:} Let us decompose the functional $E(T)$ in \eqref{ht_5x} into the sum of bilinear and linear parts
\begin{eqnarray}\label{e7.13b}
 \begin{array}{c}
E(T)=E^{(1)}(T)+E^{(2)}(T)\\[3mm]
E^{(1)}(T):=\dst\frac12\dst\int_{\Omega^\ve}\langle(\cD_{\Omega^\ve}T)(x),
     (\cD_{\Omega^\ve}T)(x)\rangle dx,\\[3mm]
E^{(2)}(T):=\dst\int_{\Omega^\ve}f(x)T(x)dx.
 \end{array}
\end{eqnarray}

The quadratic function $F(x)=x^2$ is strictly convex $[tx_1+(1-t)x_2]^2<tx^2_1+(1-t)x^2_2$ for all $x_1,x_2\in\bR$, $x_1 \neq x_2$,  $0<t<1$ and, therefore, the functional $E^{(1)}(T)$ is strictly convex. Since $E^{(2)}(T)$ is linear, the sum $E(T)=E^{(1)}(T)+E^{(2)}(T))$ is, obviously, strictly convex (see the first inequality in \eqref{ht_121}).

Next let us prove the second two-sided estimate in \eqref{ht_121}. To this end note, that the first functional
\begin{equation}\label{e7.22}
E^{(1)}(T)=\dst\frac12\dst\int_{\Omega^\ve}|(\cD_{\Omega^\ve} T)(x)|^2dx
\end{equation}
is quadratic itself. We will prove the following estimate for the second functional
\begin{equation}\label{ht_7.13c}
|E^{(2)}(T)|\leqslant M +M \left(\dst\int_{\Omega^\ve}\left|(\cD_{\Omega^\ve}T)(x)\right|^2
    dx\right)^{1/2}.
\end{equation}
Since $f\in\mathbb{L}_2(\Omega^1)$ and $T\in\wt\bH^1(\Omega^{\ve},\pa\cC\times(-\ve,\ve)) \subset\mathbb{L}_2(\Omega^\ve)$, due to Lemma \ref{l7.1} we can write
 \[
\int_{\Omega^\ve}f(x)T(x)dx\leqslant \|f|\mathbb{L}_2(\Omega^\ve)\|\
     \|T|\mathbb{L}_2(\Omega^\ve)\|\leqslant M
     \|\nabla T|\mathbb{L}_2(\Omega^\ve)\|\leqslant M\Big(1+
     \|\nabla T|\mathbb{L}_2(\Omega^\ve)\|^2\Big).
 \]
The proved inequalities justify the estimate \eqref{ht_7.13c}.

The right inequality in the second line of \eqref{ht_121} is a direct consequence of \eqref{e7.13b}, \eqref{e7.22} and \eqref{ht_7.13c}.

Let us prove the left inequality in the second line of \eqref{e7.13b}. We have
\begin{eqnarray*}
|E^{(2)}(T)|\leqslant \|f|\mathbb{L}_2(\Omega^\ve)\|\|T|\mathbb{L}_2(\Omega^\ve)\|
      \leqslant\|f|\mathbb{L}_2(\Omega^\ve)\|\|\nabla T|\mathbb{L}_2(\Omega^\ve)\|\\[2mm]
      \leqslant\frac1{2\eta}\|f|\mathbb{L}_2(\Omega^\ve)\|^2
      +\frac\eta2\|\nabla T|\mathbb{L}_2(\Omega^\ve)\|^2\leqslant\frac1{2\eta}\|f|\mathbb{L}_2(\Omega^1)\|^2
      +\frac\eta2\|\nabla T|\mathbb{L}_2(\Omega^\ve)\|^2
\end{eqnarray*}
for any $\eta>0$. Choosing $\eta<1$ and by taking $C_1=\displaystyle\frac{1-\eta}2$,
 \[
C_2\geqslant\frac1{2\eta}\|f|\mathbb{L}_2(\Omega^1)\|^2
 \]
we get
 \[
E(T)\geqslant E^{(1)}(T)-|E^{(2)}(T)|\geqslant C_1\|\nabla T|\mathbb{L}_2(\Omega^\ve)\|^2   -   C_2.
 \]
\vskip-9mm\QED
\vskip7mm

Now we perform the scaling of the variable  $t=\varepsilon \tau$, $-1<\tau<1$ and study the functionals in the fixed domain $\Omega^1=\cC \times (-1,1)$
\begin{eqnarray}\label{ht_23}
E_{\varepsilon}\left(T_{\ve}\right)&=&\int_{\Omega^1}\cK\left( \cD_{1} T_{\ve},\,\cD_{2}T_{\ve},\,\cD_{3}T_{\ve},\,\frac{1}{\varepsilon}
\cD_{4}T_{\ve},T_{\ve}\right) dx \nonumber\\
&=& \int_{-1}^{1}\int_{\cC}\cK\left(\cD_{\cC} T_{\ve},\,\frac{1}{\varepsilon}\pa_t T_{\ve},T_{\ve}\right)d\sigma dt,
\end{eqnarray}
where $\cD_{\cC}=( \cD_{1},\,\cD_{2},\,\cD_{3}),\;\cD_{4}=\pa_{t}$. The functionals $E_{\varepsilon}\left(T_{\ve}\right)$ are related to the original functional $E\left(T\right)$ by the equality
\begin{equation}\label{ht_24}
E_{\varepsilon }\left( T_{\ve}\right) =\frac{1}{\varepsilon }E
     \left(T\right),\text{ \ \ \ where \ \ \ }T_{\ve}(x,t)
     =T\left(\cx_1,\cx_2,\cx_3,\varepsilon t\right).
\end{equation}
Assume, that $T_j\in\mathbb{H}^1\left(\Omega^1\right), \; j\in\mathbb{N},$ are the scaled solutions to the problem \eqref{ht_17}, with $\ve=\ve_j, \; f_j(\cx,t)=f(\cx,\ve_j t),\; 0<\ve_j\leq 1, \; \lim\limits_{j \to \infty} \ve_j =0$;  then from the Euler-Lagrange equation, associated with the functional (see \eqref{ht_7}), follows that
\begin{eqnarray}\label{ht_25}
E_{\ve_j}(T_j):&\hskip-3mm=&\hskip-3mm\int_{\Omega^1}\cK\left( \cD_{\cC} T_j,\,\frac{1}{\varepsilon_j}\pa_tT_j,T_j\right) dx \\
&\hskip-3mm=&\hskip-3mm\int\limits_{-1}^1\int\limits_{\cC}\Big[\frac12
     \left[\left|(\cD_{\cC}T_j)(\cx,t)\right|^2+\frac1{\varepsilon^2_j}\left|(\pa_tT_j)(\cx,t)
     \right|^2\right]+f_j(\cx,t)T_j(\cx,t)\Big]d\sigma dt=0.  \nonumber
 \end{eqnarray}
From \eqref{Poinc}, \eqref{ht_25}, \eqref{ht_5x} and Lemma \ref{l7.2} follows
\begin{eqnarray}\label{ht_26}
 C_0\|T_j|\mathbb{H}^1(\Omega^1)\|^2&\hskip-3mm\leqslant&\hskip-3mm
     \|\nabla_{\Omega^1} T_j|\mathbb{L}_2(\Omega^1)\|^2=\int_{\Omega^1}
     \left[\frac{1}{2}\left\langle\cD_{\cC} T_j,\cD_{\cC} T_j\right\rangle+
     \frac1{2\ve^2_j}|\pa_t T_j|^2\right]dx\nonumber\\
&\hskip-3mm=&\hskip-3mm\left|\int\limits_{-1}^1\int\limits_{\cC}f_j(\cx,t)T_j(\cx,t)
     d\sigma\,dt\right|\leqslant\|f_j|\mathbb{L}_2(\Omega^1)\|\|T_j|\mathbb{L}_2(\Omega^1)\|
     \nonumber\\[2mm]
&\hskip-3mm\leqslant&\hskip-3mm\|f_j|\mathbb{L}_2(\Omega^1)\|\|T_j|\mathbb{H}^1(\Omega^1)\|
\end{eqnarray}
and, consequently,
\begin{eqnarray}\label{e7.25}
\left(\int_{\Omega^1}\left(\frac12\left\langle
     \cD_{\cC} T_j,\cD_{\cC} T_j\right\rangle+\frac1{2\ve^2_j}|\pa_t T_j|^2\right)
     dx\right)^{1/2}\leqslant\|T_j|\mathbb{H}^1(\Omega^1)\|\nonumber\\
\leqslant\frac1{C_0}\|f_j|\mathbb{L}_2(\Omega^1)\|\leqslant\frac2{C_0}
     \|f|\mathbb{L}_2(\Omega^1)\|,
\end{eqnarray}
for all $\ve_j<\ve_0$ and some $\ve_0>0$, because (cf. \eqref{e7.7})
\begin{eqnarray*}
\|f_j|\mathbb{L}_2(\Omega^1)\|=\left[\int_{\cC}\int_{-1}^1|f(\cx,\ve_jt)|^2d\sigma dt
     \right]^{1/2}=\left[\frac1{\varepsilon_j}\int_{-\ve_j}^{\ve_j}\int_{\cC}|f(\cx,\tau)|^2
     d\sigma d\tau\right]^{1/2}\\
\leqslant 2\|f(\cdot,0)|\mathbb{L}_2(\cC)\|,\qquad \ve_j<\ve_0.
\end{eqnarray*}

From \eqref{e7.25} follows
\begin{equation}\label{ht_26.1}
\sup_{j}\int_{\Omega^1 }\left\vert T_{j}\right\vert
^{2}dx<\infty, \quad \sup_{j}\int_{\Omega^1 }\left\vert \cD_{\cC} T_{j}\right\vert
^{2}dx<\infty ,\quad  \sup_{j}\text{\ }\frac{1}{
\varepsilon _{j}^{2}}\int_{\Omega^1 }\left\vert \pa_t T_{j}\right\vert ^{2}dx < \infty.
\end{equation}

Due to \eqref{ht_26.1} and Lemma \ref{l7.1} the sequence $\left\{T_j\right\}_{j=1}^\infty$ is uniformly bounded in  $\mathbb{H}^1\left(\Omega^1\right)$ and a weakly converging subsequence (say $\left\{T_j\right\}_{j=1}^\infty$ itself) to a function $T$ in $\mathbb{H}^1\left(\Omega^1\right)$ can be extracted.

The functional
$$
E_3(T)=\int\limits_{\Omega^1}\left\vert \pa_t T\right\vert ^{2}dx
$$
is convex and continuous in $\mathbb{H}^1\left(\Omega^1 \right)$; then it is weakly lower semi-continuous and $\pa_t T=0$ a.e., because
\begin{equation}\label{e7.28}
\int_{\Omega^1 }\left\vert \pa_t T\right\vert ^{2}dx =E_3(T)\leq\lim_j \inf E_3(T_j)
    =\lim_j\inf\int_{\Omega^1}\left|\pa_tT_{j}\right|^2dx=0
\end{equation}
(see the last inequality in \eqref{ht_26.1}). Hence $T(\cx,t)$ is independent of $t$, i.e.
\begin{equation}\label{ht_26a}
T(\cx,t)=T(\cx), \qquad \cx \in \cC, \quad -1\leq t \leq 1.
\end{equation}
Let $f_j(\cx,t):=f(\cx,\ve_jt)\to f(\cx,0)$ in $\mathbb{L}^2(\Omega^1)$. Set
\begin{eqnarray}\label{ht_26b}
 \begin{array}{c}
E^{(0)}(T)=E^{(1)}(T)+E^{(2)}(T)\\[3mm]
E^{(1)}(T):=\dst\frac12\dst\int_{\Omega^\ve}\langle(\cD_{\Omega^\ve}T)(\cx),
     (\cD_{\Omega^\ve}T)(\cx)\rangle d\cx\,dt\\
\hskip30mm=\dst\int_\cC\langle(\cD_{\Omega^\ve}T)(\cx),
     (\cD_{\Omega^\ve}T)(\cx)\rangle d\cx,\\[3mm]
E^{(2)}(T):=\dst\int_{\Omega^\ve}f(\cx,0)T(\cx)d\cx\,dt
     =2\dst\int_{\cC}f(\cx,0)T(\cx)d\cx.
 \end{array}
\end{eqnarray}

Let us check that the $E_{j}$ sequence $\Gamma$-convergs to $E^{(0)}$ in $\mathbb{H}^{1}\left( \Omega^1 \right)$. Indeed, if $T_{j}\rightharpoonup T$ in $\mathbb{H}^1\left(\Omega^1\right) $,  We have
$$
E_j(T)=E_j^{(1)}(T)+E_j^{(2)}(T),
$$
where
$$
E_j^{(1)}(T)=\int\limits_{\Omega^1 }\left( \frac{1}{2}\left\langle\cD_{\cC} T,\cD_{\cC} T\right\rangle +\frac{1}{\ve_j^2}|\pa_t T|^2 \right)dx,\quad E_j^{(2)}(T)=\int\limits_{\Omega^1 } f_jT dx.
$$
The functional  $E^{(1)}(T)$  is convex and continuous and so it is weakly lower semicontinuous  in $\mathbb{H}^{1}\left( \Omega^1 \right)$, therefore
$$
\lim \inf_j E_j^{(1)}(T_j)\geqslant \lim \inf_j E^{(1)}(T_j) \geqslant E^{(1)}(T).
$$
Sequence $E_j^{(2)}(T_j)$ converges to $E^{(2)}(T)$, because $f_j(\cx,t) \rightarrow f(\cx,0)$ and  $T_j\rightharpoonup T$ in $\mathbb{L}^2(\Omega^1)$. Consequently
$$
\lim \inf_j E_j (T_j)\geqslant E^{(0)}(T).
$$
This proves $\lim \inf$ inequality for the sequence $E_j$.

Note, that
\begin{equation}
\label{ht_27}
E^{(2)}(T)=\int\limits_{\cC}\int\limits_{-1}^1 f(\cx,0) T(\cx) dt\,
     d\sigma=2\int\limits_{\cC}f(\cx,0)T(\cx)d\sigma.
\end{equation}
To show that the lower bound is reached i.e. to build a recovery sequence $T_j$ we fix $T\in\mathbb{H}^{1}\left( \cC \right) $ and set $T\left( \cx,t\right) =T \left(\cx\right), \; \cx\in \cC,\; t \in (-1,1) $.
Define recovery sequence as $T_j(x,t)=T(x,t)=T(x)$
Then $\partial_t T_j=\partial_t T=0$ and
$$
\lim_{j\to\infty}E_j(T_j)=\lim_{j\to\infty}E_j^{(1)}(T)
+\lim_{j\to\infty}E_j^{(2)}(T)=E^{(1)}(T)+E^{(2)}(T)=E^{(0)}(T).
$$
We have proved the following result.
 %
\begin{theorem}\label{t7.3}
If $\varepsilon \rightarrow 0$ and $f_{\varepsilon}(\cx,t):=f(\cx,\varepsilon t)\to f(\cx,0)$ in $\mathbb{H}^{-1}(\Omega^1)$, then the functional in \eqref{ht_23} $\Gamma$-converges to the functional
\begin{eqnarray}\label{ht_28}
E^{(0)}(T)=\int\limits_{-1}^1\int\limits_{\cC}\left[ \frac{1}{2}\left\langle\cD_{\cC}
     T(\cx),\cD_{\cC} T(\cx)\right\rangle +f(\cx,0)T(\cx)\right]
     d\sigma\,dt\nonumber\\
=2\int\limits_{\cC}\left[\frac{1}{2}\left\langle\cD_{\cC} T(\cx),\cD_{\cC}
     T(\cx)\right\rangle + f(\cx,0)T(\cx)\right]d\sigma.
\end{eqnarray}

The following Dirichlet boundary value problem for Laplace-Beltrami equation on the mid surface $\cC$
\begin{eqnarray}\label{e7.cc}
\begin{array}{l}
\Delta_\cC T(\cx)=f(\cx,0) \quad \cx \in \cC,\\[3mm]
T^+(\cx)=0, \qquad \cx \in \pa\cC
\end{array}
 \end{eqnarray}
is an equivalent reformulation of the minimization problem with the energy functional \eqref{ht_28} (see Theorem \ref{t6.1}) and, therefore, can be considered as the $\Gamma$-limit of the initial BVP
\end{theorem}

Now we ar able to prove the main Theorem \ref{t0.1} formulated in the introduction.

\noindent
{\bf Proof of Theorem \ref{t0.1}}: Due to Corollary \ref{c.cor} the minimization problem for the functional \eqref{e7.38} is an equivalent reformulation of the BVP \eqref{e7.37}. Let us rewrite the scaled energy functional \eqref{e7.38a} as follows:
\begin{eqnarray*}
E_{\ve_j}(T_j)&\hskip-3mm=&\hskip-3mm\int\limits_{-1}^1\int\limits_{\cC}\Big[\frac12
     \left[\left|(\cD_{\cC}T_j)(\cx,t)\right|^2+\frac1{\ve^2_j}\left|(\pa_tT_j)(\cx,t)
     \right|^2\right]+f_j(\cx,t)T_j(\cx,t)\Big]d\sigma dt\nonumber\\
&&+\frac1{\ve_j}\int\limits_{\cC}q(\cx,\ve_j)\left[T_j(\cx,\ve_j)-T_j(\cx,-\ve_j)\right]
     d\sigma\nonumber\\
&&+\frac1{\ve_j}\int\limits_{\cC}T_j(\cx,-\ve_j)\left[q(\cx,\ve_j) -
     q(\cx,-\ve_j)\right]d\sigma\nonumber\\
 \end{eqnarray*}
\begin{eqnarray}\label{e7.43}
\hskip15mm&\hskip-3mm=&\hskip-3mm\int\limits_{-1}^1\int\limits_{\cC}\Big[\frac12
     \left[\left|(\cD_{\cC}T_j)(\cx,t)\right|^2+\frac1{\ve^2_j}\left|(\pa_tT_j)(\cx,t)
     \right|^2\right]+f_j(\cx,t)T_j(\cx,t)\Big]d\sigma dt\nonumber\\
&&+\frac1{\ve_j}\int\limits_{\cC}\int\limits_{-\ve_j}^{\ve_j}q(\cx,\ve_j)(\pa_tT_j)(\cx,t)
    d\sigma\,dt\nonumber\\
&&+\frac1{\ve_j}\int\limits_{\cC}T_j(\cx,-\ve_j)\left[q(\cx,\ve_j) -
     q(\cx,-\ve_j)\right]d\sigma.
 \end{eqnarray}
Since $(\pa_tT_j)(\cx,t)$ converges weakly to $\pa_tT(\cx)\equiv0$ as $j\to\infty$, $\dst\frac1{\ve_j}\pa_tT_j$ is uniformly bounded in $\bL_2(\cC,\times(-1,1))$  (see \eqref{e7.28}), $q(\cdot,\ve_j)$ is uniformly bounded in $\bL_2(\cC)$, (see \eqref{ht_26.1}), Corollary \ref{c7.8} of the Lebesgue Differentiation Theorem applies and we get:
\begin{eqnarray*}
&&\frac1{\ve_j}\int\limits_{\cC}\int\limits_{-\ve_j}^{\ve_j}q(\cx,\ve_j)(\pa_tT_j)(\cx,t)
    d\sigma\,dt\\
&&=\frac1{\ve_j}\int\limits_{\cC}\int\limits_{-\ve_j}^{\ve_j}q(\cx,\ve_j)
    \left[(\pa_tT_j)(\cx,t)-(\pa_tT)(\cx)\right]d\sigma\,dt=\co(\ve_j).
 \end{eqnarray*}
Now we can continue \eqref{e7.43} as follows:
\begin{eqnarray}\label{e7.44}
E_{\ve_j}(T_j):&\hskip-3mm=&\hskip-3mm\int\limits_{-1}^1\int\limits_{\cC}\Big[\frac12
     \left[\left|(\cD_{\cC}T_j)(\cx,t)\right|^2+\frac1{\ve^2_j}\left|(\pa_tT_j)(\cx,t)
     \right|^2\right]+f_j(\cx,t)T_j(\cx,t)\Big]d\sigma dt\nonumber\\
&&+2\int\limits_{\cC}T_j(\cx)q^0(\cx)d\sigma + \co(\ve_j).
 \end{eqnarray}
From \eqref{e7.39} and \eqref{e7.44} we get finally
\begin{eqnarray*}
\lim_{\ve_j\to0}E_{\ve_j}(T_j)&\hskip-3mm=&\hskip-3mm\int\limits_{-1}^1\int\limits_{\cC}
     \Big[\frac12\left|(\cD_{\cC}T)(\cx)\right|^2+\left[f(\cx,0)+q^0(\cx)\right]
     T(\cx)\Big]d\sigma dt\\
&\hskip-3mm=&\hskip-3mm2\int\limits_{\cC}
     \frac12\Big[\left|(\cD_{\cC}T)(\cx)\right|^2+\left[f^0(\cx)+q^0(\cx)\right]
     T(\cx)\Big]d\sigma dt
\end{eqnarray*}
and \eqref{e7.40} is proved.

The concluding assertion, that the BVP \eqref{e7.42} is an equivalent reformulation of the minimization problem with the energy functional \eqref{e7.40}, is explained in Theorem \ref{t6.1}).      \QED
 %
\begin{remark}\label{r7.6}
If we take non-zero Dirichlet data in the BVP \eqref{ht_17}, we can not reformulate the BVP into the variational form \eqref{ht_5} (see Theorem \ref{t6.1}), because $T^+\not\in\wt\bH^{1/2}(\cC_N)$ while $q\in\bH^{-1/2}(\cC_N)$ and the existence of the integral in the last summand of the functional $\Phi(T)$ can be ensured only for $q\in\wt\bH^{-1/2}(\cC_N)$. Moreover, in the functional $\Phi (T)$ will emerge a new summand
 \[
\int\limits_{\cC_D}(\pa_tT)(y)g(y)d\sigma
 \]
and to ensure its existence we have to impose even more constraint on the data $q\in\wt\bH^{1/2}(\cC_N)$.
 \end{remark}

\vskip10mm
\noindent
{\bf T. Buchukuri}, {\em A.Razmadze Mathematical Institute,  Tbilisi State University, Tamarashvili str. 6, Tbilisi 0177, Georgia}\\
{\sf email: t\_buchukuri@yahoo.com}\\
\\
{\bf R. Duduchava}, {\em A.Razmadze Mathematical Institute,  Tbilisi State University, Tamarashvili str. 6, Tbilisi 0177, Georgia}\\
{\sf email: RolDud@gmail.com}\\
\\
{\bf G. Tephnadze}, {\em Department of Mathematics, Faculty of Exact and
Natural Sciences, Tbilisi State University, Chavchavadze str. 1, Tbilisi
0128, Georgia} \\
{\sf email: giorgitephnadze@gmail.com}\\
\end{document}